\documentclass[%
preprint,
 amsmath,amssymb,
 aps,
 pre,
]{revtex4-2}
\usepackage{graphicx}
\usepackage{dcolumn}
\usepackage{placeins}
\usepackage{bm}
\usepackage{physics2}
\usephysicsmodule{ab}
\usepackage{fixdif, derivative}
\newcommand{\Tr}{\operatorname{Tr}}
\newcommand{\E}{\mathbb{E}}

\begin{document}


\title{Analysis of the Hopfield Model Incorporating the
Effects of Unlearning}

\author{Shuta Takeuchi}
\email{takesyu1928@g.ecc.u-tokyo.ac.jp}
 \affiliation{Department of Physics, The University of Tokyo, 7-3-1 Hongo, Tokyo 113-0033, Japan}
\author{Takashi Takahashi}
 \email{takashi-takahashi@g.ecc.u-tokyo.ac.jp}
\affiliation{
 Institute for Physics of Intelligence, The University of Tokyo, 7-3-1 Hongo, Tokyo 113-0033, Japan
}
\author{Yoshiyuki Kabashima}
\email{kaba@phys.s.u-tokyo.ac.jp}
\affiliation{
 Institute for Physics of Intelligence, The University of Tokyo, 7-3-1 Hongo, Tokyo 113-0033, Japan
}
\affiliation{Trans-Scale Quantum Science Institute, The University of Tokyo, 7-3-1 Hongo, Tokyo 113-0033, Japan}

\date{\today}

\begin{abstract}
We analyze a variant of the Hopfield model that incorporates an unlearning mechanism based
on spin correlations in the high-temperature regime.
In particular, we focus on the effective interaction obtained from a high-temperature approximation.
In the large system limit where extensively many patterns are stored, we employ the replica method under the replica-symmetric (RS) ansatz to
characterize the model analytically. Our analysis provides a systematic and self-consistent framework
that yields order-parameter equations and stability conditions at finite temperatures for this effective model.
Within the approximation, the resulting theory provides RS estimates of the behavior of the
signal-to-noise ratio, the memory capacity, and the criteria for selecting optimal hyperparameters,
in qualitative agreement with the findings of Nokura (1996 \textit{J. Phys. A: Math. Gen.} \textbf{29} 3871). Moreover, the theoretical predictions
are compared with synchronous dynamics simulations initialized from the target stored pattern, suggesting that unlearning suppresses spurious memories and thereby improves retrieval performance, consistent with the RS analysis.
\end{abstract}

\maketitle

\section{Introduction}
\label{sec:Introduction}
The Hopfield model was originally proposed as a paradigmatic model of associative memory in neural systems~\cite{hopfield1982neural}.
By storing patterns through Hebbian learning, the model enables retrieval via energy minimization in a fully connected network.
In the limit where the numbers of neurons $N$ and stored patterns $P$ tend to infinity keeping the pattern load $\alpha = P/N$ finite,
the phase structure of the Hopfield model is well understood within the framework of spin-glass theory~\cite{amit1985storing},
exhibiting retrieval, spin-glass, and paramagnetic phases.
A critical pattern load $\alpha_c \simeq 0.14$ separates the retrieval phase from the spin-glass phase, beyond which reliable recall of stored patterns becomes impossible.

Even below this critical load, however, the energy landscape of the Hopfield model contains many locally stable states that do not correspond to any single stored pattern.
These so-called spurious memories arise from the coexistence of multiple metastable states and can significantly degrade retrieval performance~\cite{amit1985storing}.
To address this issue, Hopfield and his collaborators introduced the concept of \emph{unlearning}, in which interactions associated with spurious states are selectively weakened~\cite{hopfield1983unlearning}.
In the original formulation, the network is repeatedly initialized from random configurations and allowed to relax to fixed points,
and the resulting spin configurations are used to modify synaptic couplings via an anti-Hebbian rule.
This idea is biologically motivated by the hypothesis that REM sleep serves to eliminate unnecessary or parasitic memories~\cite{crick1983function}.

While the original unlearning scheme relies on information extracted from converged states, its essential purpose is to suppress spurious correlations in the network.
Motivated by this observation, several alternative unlearning strategies have been proposed that do not explicitly depend on convergence dynamics.
These include pseudo-inverse-type methods~\cite{fachechi2019dreaming,agliari2019dreaming}, eigenvector-based approaches~\cite{benedetti2024eigenvector},
and schemes based on spin correlations evaluated in the high-temperature phase ~\cite{nokura1996unlearning}.
More recently, related unlearning and dreaming mechanisms have been studied in Hopfield-like networks from the viewpoints of optimal retrieval, regularization, early stopping, and generalization in correlated data~\cite{benedetti2022supervised,benedetti2024training,agliari2024regularization,serricchio2025daydreaming}.
These studies have shown that unlearning is an effective
mechanism for reshaping the energy landscape, suppressing spurious memories,
and improving retrieval performance.

In the present paper, we focus particularly on the model proposed in ~\cite{nokura1996unlearning}, which we call ``$J^\prime$ model''. In this model, unlearning is implemented by modifying the Hebbian couplings
using spin correlation functions computed in the high-temperature phase.
Importantly, the inverse temperature used to evaluate these correlations is
taken to be small, $0< \gamma \ll 1$.
Since the exact evaluation of these correlations is generally difficult, we use a high-temperature approximation similar to the original formulation and analyze the resulting effective interaction.
Previous studies based on signal-to-noise (SN) analysis and numerical simulations have suggested that this mechanism enhances retrieval robustness and can increase the memory capacity ~\cite{nokura1996unlearning}.
However, these analyses were primarily qualitative and restricted to the zero-temperature limit.
As a consequence, the precise location and nature of the phase boundaries, the dependence of memory capacity on unlearning parameters,
and the behavior at finite temperatures have remained unclear.

A quantitative understanding of these issues is essential for evaluating the effectiveness and robustness of unlearning.
In particular, the dependence of the memory capacity on the unlearning strength and temperature is crucial for identifying optimal parameter choices,
while finite-temperature analysis is indispensable for understanding the stability of retrieval under noise and thermal fluctuations.

The purpose of this paper is to provide a systematic statistical-mechanics analysis of the effective $J^\prime$ model and to clarify the effects of unlearning within a unified theoretical framework.
Using the replica method under the replica symmetric (RS) ansatz, we derive the free energy and the corresponding saddle-point equations in the large system limit of $N \to \infty$ with finite pattern load $\alpha$.
Within this RS framework, we analyze the retrieval performance, phase diagram,
and stability conditions at finite temperatures, and compare the theoretical
predictions with numerical simulations based on synchronous dynamics.
Furthermore, we evaluate the signal-to-noise ratio within the replica framework, examine consistency with previous qualitative studies,
and estimate the memory capacity and its dependence on unlearning parameters theoretically.

This paper is organized as follows.
In Sec.~\ref{sec:model}, we introduce the $J^\prime$ model and derive an effective interaction by rewriting the spin-correlation term using the high-temperature approximation.
In Sec.~\ref{sec:analysis}, we perform a replica analysis under the RS ansatz and obtain the saddle-point equations.
In Sec.~\ref{sec:results}, we analyze the macroscopic properties of the system, including retrieval performance, signal-to-noise ratio, phase diagrams, and memory capacity.
Section~\ref{sec:conclusion} concludes the paper with a summary and discussion of future directions.

\section{Model Setup}
\label{sec:model}
In this section, we introduce the $J'$ model, and derive an effective form of the interaction by rewriting the spin correlation function using a high-temperature approximation.
\subsection{$J'$ model}
Let us suppose the same setup as in the Hopfield model for the spins and patterns.
Namely, given $P$ binary vectors $\bm{\xi}^1, \ldots, \bm{\xi}^P
\in \{+1, -1\}^N$ sampled independently and uniformly, we define couplings by Hebb's rule as
\begin{align}
\label{eq:Hebb-rule}
J_{ij} = \frac{1}{N} \sum_{\mu=1}^P (1-\delta_{ij}) \xi_i^\mu \xi_j^\mu .
\end{align} 
The $J'$ model incorporates unlearning by using the spin correlation function
computed in the high-temperature phase of the Hopfield model, as
\begin{align}
\label{eq:Jprime-interaction}
    J'_{ij} =
    \begin{cases}
        J_{ij} - \epsilon \left< S_i S_j \right>_\gamma & (i \neq j), \\
        0 & (i = j).
    \end{cases}
\end{align}
Here, $\epsilon$ is the unlearning strength and $\gamma$ is the inverse temperature
in the high-temperature phase $(0 < \gamma \ll 1)$, and $J_{ij}$ is defined by standard Hebbian rule.
The use of the high-temperature spin correlation function is motivated by the
expectation that it contains rich information on spurious memories, while being only
weakly affected by the retrieval states.
With this interaction $J'$, the model is defined by
\begin{align}
    &H(\bm{S}) = -\frac{1}{2}\sum_{i\neq j} J'_{ij} S_i S_j 
    = -\frac{1}{2}\bm{S}^\mathsf{T} J' \bm{S}, 
    \qquad \bm{S}=(S_1,\ldots,S_N)^\mathsf{T}, \\
    &P(\bm{S}|\{\bm{\xi}^\mu\}) 
    = \frac{\exp\ab(-\beta H(\bm{S}))}{\Tr \exp\ab(-\beta H(\bm{S}))},
\end{align}
where $\mathrm{Tr}$ denotes 
the summation or integration over all possible states. 
The spin correlation function is computed as the thermal average at inverse temperature $\gamma$:
\begin{align}
    \left<S_i S_j \right>_\gamma &= \frac{1}{Z_\gamma}\Tr S_i S_j \exp \left(\frac{\gamma}{2}\sum_{i\neq j}J_{ij}S_i S_j\right) \notag \\
    Z_\gamma &= \Tr \exp \left(\frac{\gamma}{2}\sum_{i\neq j}J_{ij}S_i S_j\right).
\end{align}

\subsection{Rewriting Interactions by High-Temperature Expansion}
As it stands in Eq.~(\ref{eq:Jprime-interaction}), it is computationally difficult to evaluate the correlation function $\left \langle S_i S_j \right \rangle_\gamma$.
As a practical solution, using the assumption $0<\gamma\ll1$ in the original model, we estimate it by the high-temperature approximation.
In the high-temperature phase, high-temperature expansion yields
\begin{align}
    \left<S_i S_j \right>_\gamma = \delta_{ij} + \gamma \sum_k J_{ik} \langle S_k S_j\rangle_\gamma + \cdots
\end{align}
for sufficiently small positive $\gamma$. Neglecting $\mathcal{O}(\gamma^2)$ terms in this expression yields the approximation: 
\begin{align}
\left \langle S_i S_j \right \rangle_{\gamma} \simeq \left[ (I-\gamma J)^{-1} \right]_{ij}, 
\end{align} 
where $I$ denotes the identity matrix.
Using this approximation, the interaction of the $J'$ model is redefined as
\begin{align}
    \label{eq:interaction-Jprime}
    J' = J - \epsilon (I - \gamma J)^{-1}.
\end{align}
In this paper, we analyze the effective model defined by
Eq.~(\ref{eq:interaction-Jprime}), rather than the original correlation-based
unlearning rule itself.

The replacement of the spin correlation function by
$(I-\gamma J)^{-1}$ is used here as the high-temperature correlation
approximation introduced in the original formulation.
Thus, the results below should be interpreted within the regime where this
approximation gives a meaningful effective interaction.
A systematic assessment of its quantitative accuracy beyond the
high-temperature regime is left for future work.

For the effective interaction in Eq.~(\ref{eq:interaction-Jprime}) to be well
defined, the matrix $I-\gamma J$ must be invertible and positive definite.
Let $\{\lambda_k\}_{k=1}^N$ be the eigenvalues of $J$.
Then this condition is equivalent to
$1-\gamma\lambda_k > 0$ for all $k=1,\ldots,N$.
This is equivalent to the convergence condition of the Neumann expansion, which requires $\max_{1\le k\le N} |\gamma \lambda_k| < 1$. 
In the present setting, $\gamma$ is introduced as a sufficiently small inverse temperature
associated with the high-temperature correlation approximation, and the pattern
load $\alpha$ is not taken to be very large.
For typical Hopfield couplings in the parameter regimes considered here, the
spectrum of $\gamma J$ is expected to remain away from unity, so that
$I-\gamma J$ generically satisfies the condition.
Nevertheless, if an eigenvalue approaches $1/\gamma$, the inverse term can be
strongly enhanced, and the corresponding parameter region should be interpreted
with caution.

In applying Eq.~(\ref{eq:interaction-Jprime}) to the Ising Hamiltonian, the
diagonal terms of $J'$ give only configuration-independent contributions,
because $S_i^2=1$.
Below they are therefore omitted in the replica calculation, consistently with the
definition $J'_{ii}=0$ in Eq.~(\ref{eq:Jprime-interaction}).
With this convention, the limiting cases are consistent with the standard
Hopfield model.
When the unlearning strength vanishes, $\epsilon\to0$, one immediately obtains
$J'\to J$.
In the limit $\gamma\to0$, one has $(I-\gamma J)^{-1}\to I$, and hence
$J'\to J-\epsilon I$.
The second term is diagonal and is removed by the above convention.
Thus the off-diagonal couplings satisfy $J'_{ij}=J_{ij}$ for $i\neq j$, and
the model reduces to the standard Hopfield model up to this constant
contribution.

Table~\ref{tab:params} summarizes the parameters used in this study.
\begin{table}[t]
\caption{Model parameters used in this study}
\label{tab:params}
\begin{ruledtabular}
\begin{tabular}{c l l}
Symbol & Name & Definition/Role \\
\hline
$\alpha$ & Pattern load & Ratio of the number of stored patterns $P$ to neurons $N$ \\
$\epsilon$ & Unlearning strength & Strength that weakens Hebbian couplings \\
$\gamma$ & Unlearning inverse temperature & Sets $\langle S_i S_j\rangle_\gamma
\approx (I-\gamma J)^{-1}_{ij}$ \\
$\beta$ & Physical inverse temperature & Temperature of the analyzed model \\
\end{tabular}
\end{ruledtabular}
\end{table}

\section{Replica analysis}
\label{sec:analysis}
In this section, we analyze the $J'$ model using the replica method. 
For this purpose, we first evaluate the $n$-th moment of the partition function for $n\in \mathbb{N}$  introducing auxiliary variables to handle the matrix inversion. Then, we analytically continue the obtained expression of the moment  of the partition function to real numbers $n \in \mathbb{R}$ under the RS ansatz, which yields the average free energy. Finally, we take the zero temperature limit for examining the property of the local minima of the energy function of Eq.~\eqref{eq:Jprime-interaction}. 
\subsection{Handling matrix inversion}
The partition function to be analyzed is
\begin{align}
    Z = \Tr \exp \left(\frac{\beta}{2}\bm{S}^{\mathsf{T}}J'\bm{S} \right) = \Tr \exp \left(\frac{\beta}{2}\bm{S}^\mathsf{T}J \bm{S} - \frac{\epsilon\beta}{2}\bm{S}^\mathsf{T}(I - \gamma J)^{-1}\bm{S} \right).
\end{align}
To handle the inverse of $I - \gamma J$, we introduce a new variable $\bm{\phi}$ via a Hubbard--Stratonovich transformation.
\begin{align}
    Z = \int^{i\infty}_{-i\infty} \mathrm{d} \bm{\phi} \Tr \exp \left(\frac{\beta}{2}\bm{S}^\mathsf{T}J \bm{S} + \frac{\beta}{2} \bm{\phi}^{\mathsf{T}} (I - \gamma J)\bm{\phi} + \beta\sqrt{\epsilon} \bm{S}^{\mathsf{T}}\bm{\phi}\right) \frac{\sqrt{\det(I - \gamma J)}}{(2\pi i)^{N/2}}.
\end{align}
The factor $\det(I-\gamma J)$ depends only on the quenched coupling matrix $J$
and not on the spin configuration $\bm S$ or the auxiliary variable $\bm\phi$.
It therefore gives an additive contribution to $\log Z$ that is independent of
the order parameters introduced below.
The saddle-point equations determine these order parameters and hence describe
the statistical properties of the spin configurations.
Consequently, this factor does not affect the saddle-point equations.
Moreover, since it gives only a constant shift of the free energy at fixed
$(\alpha,\epsilon,\gamma,\beta)$, it does not affect the comparison between
different saddle-point solutions or the phase diagrams discussed below.
Therefore, we omit this determinant factor and other normalization
constants, since our analysis focuses on the saddle-point equations, phase
boundaries, and order parameters.
Neglecting other constant contributions, we will analyze the following 
\begin{align}
    \label{eq:partition-func}
    \tilde{Z} = \int^{i\infty}_{-i\infty} \mathrm{d} \bm{\phi} \Tr \exp \left(\frac{\beta}{2}\bm{S}^{\mathsf{T}}J \bm{S} + \frac{\beta}{2} \bm{\phi}^{\mathsf{T}} (I - \gamma J)\bm{\phi} + \beta\sqrt{\epsilon} \bm{S}^{\mathsf{T}}\bm{\phi}\right).
\end{align}

\subsection{Replica method}
The typical properties of the system can be analyzed by evaluating the configurational average of 
$\log \tilde{Z}$ with respect to the stored patterns $\bm{\xi}^1, \ldots, \bm{\xi}^P$. However, a direct evaluation of this quantity is difficult. 

To overcome this difficulty, we employ the replica method.  Within this framework, the disorder average $\mathbb{E}[\log \tilde{Z}]$ is computed using the identity
\begin{align}
\label{eq:replica_trick}
\mathbb{E}\ab[\log \tilde{Z}]
= \lim_{n\rightarrow 0}
\frac{\mathbb{E}\ab[\tilde{Z}^n] - 1}{n},
\end{align}
which is evaluated by first computing $\mathbb{E}\ab[\tilde{Z}^n]$ for natural numbers $n \in \mathbb{N}$ and then analytically continuing the result to real values $n \in \mathbb{R}$ \cite{mezard1987spin}.

Without loss of generality, we assume that the pattern to be retrieved is $\bm{\xi}^1 = \bm{1} = \ab(1,1,\ldots,1)^\mathsf{T}$. We then evaluate $\mathbb{E}_{\Bab{\bm{\xi}^\mu}} \ab[\tilde{Z}^n]$ by separating the contribution of $\bm{\xi}^1$ from those of the remaining patterns. Since all patterns are sampled independently, their contributions factorize. This procedure leads to
\begin{align}
  \label{eq:average-of-partitionfunction}
  &\mathbb{E}_{\{\bm{\xi}^\mu\}} \ab[\tilde{Z}^n] \notag \\
  &= \Tr \int_{-i\infty}^{i\infty} \d \bm{\phi}
  \exp\ab(
  \beta\frac{1 + \alpha\gamma}{2}\sum_{i,a}(\phi_i^a)^2
  + \beta \sqrt{\epsilon}\sum_{i,a}S_i^a \phi_i^a
  + \frac{\beta}{2N}\sum_{a}\ab(\sum_i S_i^a)^2
  - \frac{\beta \gamma}{2N}\sum_{a}\ab(\sum_i\phi_i^a)^2) \notag \\
  &\quad\times
  \ab(\mathbb{E}_{\bm{\xi}}\ab[
  \exp \ab(
  \frac{\beta}{2}\sum_{a}\ab(\frac{1}{\sqrt{N}}\sum_i \xi_i S_i^a)^2
  - \frac{\beta \gamma}{2}\sum_{a}\ab(
  \frac{1}{\sqrt{N}}\sum_i \xi_i \phi_i^a)^2
  )])^{P-1},
\end{align}
where we have denoted a representative non-retrieved pattern by $\bm{\xi}$
after dropping the pattern index $\mu=2,3,\ldots,P$, since all these patterns
give identical contributions.

To perform the average over this random vector $\bm{\xi}$, we introduce the variables
\begin{align*}
    v_a = \frac{1}{\sqrt{N}}\sum_i \xi_i S_i^a,
    \qquad
    w_a = \frac{1}{\sqrt{N}}\sum_i \xi_i \phi_i^a .
\end{align*}
For fixed $\{S^a,\phi^a\}$, the central limit theorem implies that
$\{v_a,w_a\}$ follow a multivariate Gaussian distribution with zero mean.
Their covariances are determined by the overlaps
$N^{-1}\sum_i S_i^a S_i^b$,
$N^{-1}\sum_i \phi_i^a \phi_i^b$,
and
$N^{-1}\sum_i S_i^a \phi_i^b$.
Consequently, Eq.~(\ref{eq:average-of-partitionfunction}) depends on
$\bm{S}^a$ and $\bm{\phi}^a$ only through macroscopic quantities such as
$N^{-1}\sum_i S_i^a$ and $N^{-1}\sum_i S_i^a S_i^b$.
This motivates the introduction of the following order parameters:
\begin{align}
    m_a &= \frac{1}{N}\sum_i S_i^a, \\
    u_a &= \frac{1}{N}\sum_i \phi_i^a, \\
    q_{ab} &= \frac{1}{N}\sum_i S_i^a S_i^b, \\
    r_{ab} &= \frac{1}{N}\sum_i S_i^a \phi_i^b, \\
    p_{ab} &= \frac{1}{N}\sum_i \phi_i^a \phi_i^b .
\end{align}
Here, $m_a$ represents the overlap with the target pattern
$\bm{\xi}^1=\bm{1}$, and $q_{ab}$ denotes the overlap between replicas
$\bm{S}^a$ and $\bm{S}^b$.
The remaining order parameters arise from the introduction of the unlearning
mechanism and are absent in the standard Hopfield model.

We enforce these definitions by introducing conjugate variables (denoted by
hats) via the Fourier representation of the delta function.
For example,
\begin{align}
    1 &= \int \mathrm{d}q_{ab}\,
    \delta\ab(N q_{ab} - \sum_i S_i^aS_i^b) \notag \\
    \label{eq:fourier-representation-of-the-delta-function}
    &= \int \mathrm{d}q_{ab}
    \int^{i\infty}_{-i\infty}\frac{\mathrm{d} \hat{q}_{ab}}{2\pi i}
    \exp\ab(-\hat{q}_{ab}\ab(N q_{ab} - \sum_i S_i^aS_i^b)).
\end{align}
This representation allows all constraints to be expressed in exponential
form, which greatly simplifies the analysis.

In the thermodynamic limit $N,P\to\infty$ with the pattern load $\alpha=P/N=\mathcal{O}(1)$,
the disorder-averaged moment of the partition function is dominated by
contributions exponential in $N$.
This enables the use of the saddle-point method, and the configurational
average is determined by the extremum of the integrand with respect to the
order parameters.
Specifically, we obtain
\begin{align}
\lim_{N\to\infty} \frac{1}{N}\log \mathbb{E}[\tilde{Z}^n]
= -\beta n f,
\end{align}
where
\begin{align}
\label{eq:free-energy}
    -n\beta f
    &=
    \mathop{\rm extr}
    \Biggl\{
    -\beta\left(
    \sum_a (\hat{m}_a m_a + \hat{u}_a u_a)
    + \sum_{a,b}(\hat{q}_{ab}q_{ab}
    + \hat{r}_{ab}r_{ab}
    + \hat{p}_{ab}p_{ab})
    \right) \notag \\
    &\quad
    + \frac{\beta}{2}\sum_a m_a^2
    - \frac{\beta \gamma}{2}\sum_a u_a^2
    + \alpha \log \int\mathrm{d} \bm{x}\,
    \mathrm{d} \bm{y}\, e^{\beta K}
    + \log \int_{-i\infty}^{i\infty}
    \mathrm{d} \bm{\phi}\, \Tr e^{\beta L}
    \Biggr\}.
\end{align}
with
\begin{align}
  L &= \frac{1 + \alpha \gamma}{2}\sum_a (\phi^a)^2
  + \sqrt{\epsilon}\sum_a S^a \phi^a
  + \sum_a \hat{m}_a S^a
  + \sum_a \hat{u}_a \phi^a
  + \sum_{a,b}\hat{q}_{ab}S^a S^b \notag \\
  &\quad
  + \sum_{a,b}\hat{r}_{ab}S^a \phi^b
  + \sum_{a,b}\hat{p}_{ab}\phi^a \phi^b, \\
  K &= -\frac{i\gamma }{2\beta}\sum_a (x^a)^2
  + \frac{i}{2\beta}\sum_a (y^a)^2
  + \frac{i\gamma}{2}\sum_{a,b}q_{ab}x^a x^b
  + \frac{i\gamma}{2}\sum_{a,b}p_{ab}y^a y^b
  + i\gamma\sum_{a,b}r_{ab}x^a y^b .
\end{align}
Here, \(\mathop{\rm extr}\) denotes extremization with respect to the order
parameters and their conjugate variables.
The saddle-point equations follow from this extremization condition, and the
free energy is obtained by evaluating the expression at the corresponding
self-consistent solution.
Details of the derivation are provided in
Appendix~\ref{detailed-calc-replica}.

\subsection{RS ansatz}
Without further assumptions on the structure of the saddle point, it is difficult
to evaluate Eq.~\eqref{eq:replica_trick}, since
Eq.~(\ref{eq:free-energy}) still explicitly depends on the discrete replica number
$n$.
The key to resolving this difficulty is to focus on replica symmetry.
Because Eq.~(\ref{eq:free-energy}) is invariant under any relabeling or permutation
of the replica indices, it is natural to assume that all replicas are equivalent.
We therefore impose the following RS ansatz.
To simplify the calculation and to match the scaling with $\beta$, we parametrize
the solution as
\begin{align}
  q_{ab} &= 
  \begin{cases}
      q + \chi_q, & a = b, \\
      q, & a \neq b,
  \end{cases}
  \label{eq:qab} \\
  r_{ab} &= 
  \begin{cases}
      r + \chi_r, & a = b, \\
      r, & a \neq b,
  \end{cases} \\
  p_{ab} &= 
  \begin{cases}
      p + \chi_p, & a = b, \\
      p, & a \neq b,
  \end{cases} \\
  m_a &= m \quad (\forall a), 
  \qquad 
  u_a = u \quad (\forall a).
\end{align}

For the conjugate variables, we assume
\begin{align}
  \hat{q}_{ab} &=
  \begin{cases}
      \beta \hat{q} + \hat{\chi}_q, & a = b, \\
      \beta \hat{q}, & a \neq b,
  \end{cases} \\
  \hat{r}_{ab} &=
  \begin{cases}
      \beta \hat{r} + \hat{\chi}_r, & a = b, \\
      \beta \hat{r}, & a \neq b,
  \end{cases} \\
  \hat{p}_{ab} &=
  \begin{cases}
      \beta \hat{p} + \hat{\chi}_p, & a = b, \\
      \beta \hat{p}, & a \neq b,
  \end{cases} \\
  \hat{m}_a &= \hat{m}, 
  \qquad 
  \hat{u}_a = \hat{u} \quad (\forall a).
  \label{eq:mhat_uhat}
\end{align}

Substituting these expressions into Eq.~(\ref{eq:free-energy}), Eq.~(\ref{eq:replica_trick}) yields
\begin{align}
  -\beta f
  &=
  \mathop{\rm extr}
  \Biggl\{
  -\beta\Bigl(
  \hat{m}m + \hat{u}u
  + \beta \hat{q}\chi_q + \hat{\chi}_q q + \hat{\chi}_q \chi_q
  + \beta \hat{r}\chi_r + \hat{\chi}_r r + \hat{\chi}_r \chi_r
  + \beta \hat{p}\chi_p + \hat{\chi}_p p + \hat{\chi}_p \chi_p
  \Bigr) \notag \\
  &\quad
  + \frac{\beta}{2}m^2
  - \frac{\beta\gamma}{2}u^2
  - \frac{\alpha}{2}\log \Delta
  + \frac{\alpha\beta}{2\Delta}
  \Bigl(
  (1 + \beta \gamma \chi_p)q
  - \gamma(1 - \beta \chi_q)p
  - 2\beta \gamma \chi_r r
  \Bigr)
  + \beta \hat{\chi}_q \notag \\
  &\quad
  - \frac{1}{2}\log \bigl(1 + \alpha \gamma + 2\hat{\chi}_p \bigr)
  - \beta
  \frac{(\hat{\chi}_r + \sqrt{\epsilon})^2 + \hat{u}^2 + 2\hat{p}}
  {2(\alpha \gamma + 2\hat{\chi}_p + 1)} \notag \\
  &\quad
  + \int Dy \,
  \log 2\cosh \beta
  \left(
  \hat{m} - A\hat{u} + \frac{\delta}{\sqrt{2}}y
  \right)
  \Biggr\},
\end{align}
where
\begin{align}
  \Delta
  = (1 - \beta \chi_q)(1 + \beta\gamma \chi_p)
  + \beta^2 \gamma \chi_r^2 .
\end{align}
The saddle-point equations are obtained by extremizing this RS free energy with
respect to the RS order parameters and their conjugate variables.

The saddle-point equations for the order parameters are then given by
\begin{align}
  m
  &= \int Dy \,
  \tanh \beta
  \left(
  m + A\gamma u + \frac{\delta}{\sqrt{2}}y
  \right), \label{eq:saddle-equation} \\
  u
  &= -\frac{1}{\alpha \gamma + 2 \hat{\chi}_p + 1}
  \left(
  - \gamma u + (\sqrt{\epsilon}+\hat{\chi}_r)m
  \right), \\
  q
  &= \int Dy \,
  \tanh^2 \beta
  \left(
  m + A\gamma u + \frac{\delta}{\sqrt{2}}y
  \right),
  \qquad
  \chi_q = 1 - q, \\
  \chi_p
  &= -\frac{1}{\beta(\alpha\gamma + 2\hat{\chi}_p + 1)}
  + A^2 \chi_q, \\
  p
  &= \frac{1}{(\alpha \gamma + 2 \hat{\chi}_p + 1)^2}
  \Bigl(
  \gamma^2 u^2 + 2\hat{p}
  + 2(\sqrt{\epsilon}+\hat{\chi}_r)
  \bigl(-\gamma m u + \beta \chi_q(\hat{r} - 2\hat{p} A)\bigr)
  + (\sqrt{\epsilon} + \hat{\chi}_r)^2 q
  \Bigr), \\
  r
  &= -\frac{1}{\alpha \gamma + 2 \hat{\chi}_p + 1}
  \Bigl(
  -\gamma u m
  + \beta \chi_q(\hat{r} - 2\hat{p} A)
  + (\sqrt{\epsilon}+\hat{\chi}_r)q
  \Bigr),
  \qquad
  \chi_r = -A \chi_q .
\end{align}

The equations for the conjugate variables read
\begin{align}
  \hat{q}
  &= \frac{\alpha}{2\Delta^2}
  \Bigl(
  (1 + \beta \gamma \chi_p)^2 q
  - 2\beta \gamma \chi_r r (1 + \beta \gamma \chi_p)
  + \beta^2 \gamma^2 \chi_r^2 p
  \Bigr),
  \qquad
  \hat{\chi}_q
  = \frac{\alpha}{2\Delta} (1 + \beta \gamma \chi_p), \\
  \hat{p}
  &= \frac{\alpha\gamma^2}{2\Delta^2}
  \Bigl(
  \beta^2 \chi_r^2 q
  + 2\beta \chi_r r (1 - \beta \chi_q)
  + (1 - \beta \chi_q)^2 p
  \Bigr),
  \qquad
  \hat{\chi}_p
  = -\frac{\alpha\gamma}{2\Delta} (1 - \beta \chi_q), \\
  \hat{r}
  &= \frac{\alpha\gamma}{\Delta^2}
  \Bigl(
  -\beta (1 + \beta \gamma \chi_p)\chi_r q
  - (\Delta - 2\beta^2 \gamma \chi_r^2)r
  + \beta \gamma (1 - \beta \chi_q)\chi_r p
  \Bigr),
  \qquad
  \hat{\chi}_r
  = -\frac{\alpha\beta\gamma}{\Delta}\chi_r, \\
  \Delta
  &= (1 - \beta \chi_q)(1 + \beta\gamma \chi_p)
  + \beta^2 \gamma \chi_r^2, \\
  \Delta_2
  &= 4\hat{q}\hat{p} - \hat{r}^2, \\
  A
  &= \frac{\hat{\chi}_r + \sqrt{\epsilon}}
  {\alpha \gamma + 2 \hat{\chi}_p + 1}, \\
  \delta
  &= \sqrt{
  \frac{
  (2\hat{q} + \sqrt{\Delta_2} - A\hat{r})^2
  + (\hat{r} - 2A\hat{p} - A\sqrt{\Delta_2})^2
  }{
  \hat{q} + \hat{p} + \sqrt{\Delta_2}
  }} .
\end{align}
Here we have defined
\begin{align}
  Dy \equiv \frac{e^{-y^2/2}}{\sqrt{2\pi}}\,\mathrm{d}y .
\end{align}

The order parameter $m$ represents the overlap with the retrieved (target)
pattern and therefore serves as a measure of retrieval performance.
By examining the behavior of $m$ as a function of the pattern load $\alpha$,
one can determine whether memory retrieval is successful.
As a consistency check of the replica calculation, the standard Hopfield model
should be recovered in the limits $\epsilon \to 0$ or $\gamma \to 0$.
Indeed, a straightforward calculation shows that taking either of these limits
in the present RS saddle-point equations reproduces those of the standard
Hopfield model.

\subsection{Zero Temperature Expression}
The zero-temperature limit $(\beta \to \infty)$ describes the equilibrium
properties of the model in the absence of thermal fluctuations.
In this limit, the RS saddle-point equations characterize the local minima
selected by the thermodynamic analysis.
It is therefore useful to derive the saddle-point equations in the limit
$\beta \to \infty$.

As $\beta \to \infty$, one has $q \to 1$ and
$\chi_q,\chi_p,\chi_r \to 0$.
In this limit, we assume the standard scaling
$\beta \chi_q = \mathcal{O}(1)$,
$\beta \chi_p = \mathcal{O}(1)$, and
$\beta \chi_r = \mathcal{O}(1)$.
We thus introduce the rescaled variables
\begin{align}
C \equiv \beta \chi_q, \qquad
E \equiv \beta \chi_p, \qquad
F \equiv \beta \chi_r .
\end{align}
Taking the limit $\beta \to \infty$ under this scaling, the saddle-point equations
reduce to
\begin{align}
\label{eq:saddle-equation-zerotemp}
  m
  &= \mathrm{erf}\!\left(
  \frac{m + A\gamma u}{\delta}
  \right), \\
  u
  &= -\frac{1}{\alpha \gamma + 2 \hat{\chi}_p + 1}
  \left(
  -\gamma u + (\sqrt{\epsilon}+\hat{\chi}_r)m
  \right), \\
  C
  &= \frac{2}{\delta\sqrt{\pi}}
  \exp\!\left(
  -\frac{(m + A \gamma u)^2}{\delta^2}
  \right), \\
  p
  &= \frac{1}{(\alpha \gamma + 2 \hat{\chi}_p + 1)^2}
  \Bigl(
  \gamma^2 u^2 + 2\hat{p}
  + 2(\sqrt{\epsilon}+\hat{\chi}_r)
  \bigl(-\gamma u m + C(\hat{r} - 2\hat{p} A)\bigr)
  + (\sqrt{\epsilon} + \hat{\chi}_r)^2
  \Bigr), \\
  r
  &= -\frac{1}{\alpha \gamma + 2 \hat{\chi}_p + 1}
  \Bigl(
  -\gamma u m
  + C(\hat{r} - 2\hat{p} A)
  + (\sqrt{\epsilon}+\hat{\chi}_r)
  \Bigr), \\
  \hat{q}
  &= \frac{\alpha}{2\Delta^2}
  \Bigl(
  (1 + \gamma E)^2
  - 2\gamma F r (1 + \gamma E)
  + \gamma^2 F^2 p
  \Bigr),
  \qquad
  \hat{\chi}_q
  = \frac{\alpha}{2\Delta} (1 + \gamma E), \\
  \hat{p}
  &= \frac{\alpha\gamma^2}{2\Delta^2}
  \Bigl(
  F^2 + 2F r (1 - C) + (1 - C)^2 p
  \Bigr),
  \qquad
  \hat{\chi}_p
  = -\frac{\alpha\gamma}{2\Delta} (1 - C), \\
  \hat{r}
  &= \frac{\alpha\gamma}{\Delta^2}
  \Bigl(
  -(1 + \gamma E)F
  - (\Delta - 2\gamma F^2)r
  + \gamma (1 - C)F p
  \Bigr),
  \qquad
  \hat{\chi}_r
  = -\frac{\alpha \gamma}{\Delta}F, \\
  \Delta
  &= (1 - C)(1 + \gamma E) + \gamma F^2, \\
  \Delta_2
  &= 4\hat{q}\hat{p} - \hat{r}^2, \\
  A
  &= \frac{\hat{\chi}_r + \sqrt{\epsilon}}
  {\alpha \gamma + 2 \hat{\chi}_p + 1}, \\
  E
  &= -\frac{1}{\alpha\gamma + 2\hat{\chi}_p + 1}
  + \frac{(\hat{\chi}_r + \sqrt{\epsilon})^2\, C}
  {(\alpha\gamma + 2\hat{\chi}_p + 1)^2}, \\
  F
  &= -\frac{\hat{\chi}_r + \sqrt{\epsilon}}
  {\alpha\gamma + 2\hat{\chi}_p + 1}\, C, \\
  \delta
  &= \sqrt{
  \frac{
  (2\hat{q} + \sqrt{\Delta_2} - A\hat{r})^2
  + (\hat{r} - 2A\hat{p} - A\sqrt{\Delta_2})^2
  }{
  \hat{q} + \hat{p} + \sqrt{\Delta_2}
  }} .
\end{align}
Here, the error function is defined as
\begin{align}
  \mathrm{erf}(x)
  = \frac{2}{\sqrt{\pi}}
  \int_0^x \mathrm{d}t\, e^{-t^2}.
\end{align}
\section{Performance analysis}
\label{sec:results}
In this section, we quantitatively evaluate the effect of unlearning by
numerically solving the saddle-point equations derived in the previous section.
Since these equations are obtained under the RS ansatz,
the phase boundaries and memory capacities reported below should be understood as RS estimates.
Possible replica-symmetry-breaking corrections are examined through the AT
stability condition in Sec.~\ref{subsec:phase-diagram-and-AT-line}.
We use the overlap $m$, obtained from
Eqs.~(\ref{eq:saddle-equation}) and~(\ref{eq:saddle-equation-zerotemp}),
as a measure of retrieval performance.
The theoretical predictions are then compared with numerical simulations of the retrieval dynamics.

\subsection{Comparison with Numerical Simulations}
\label{subsec:comparing-to-simulation}

We compare the RS predictions with direct numerical simulations of retrieval
dynamics, focusing in particular on the relation between the locally continued
RS retrieval solution and the finite-size dynamical behavior.

In the Hopfield framework, stored patterns are expected to be dynamically
accessible to stable configurations of the network.
To study memory retrieval and to compare the analytical results with numerical
simulations, it is therefore necessary to specify a concrete retrieval dynamics.

A simple and commonly used choice in simulations of spin systems is asynchronous
dynamics, in which a single spin is aligned with its local field at each update.
For symmetric interactions, this dynamics monotonically decreases the energy and
therefore converges to a fixed point.
In the present numerical experiments, however, we adopt the following
synchronous, or parallel, update rule:
\begin{align}
\label{eq:synchronous-update}
    \bm{S}^{t} = \mathrm{sgn}\ab(J'\bm{S}^{t-1}). \quad (t=1,2,\ldots)
\end{align}
where $\bm{S}^t = (S_1^t,S_2^t,\ldots,S_N^t)^{\mathsf{T}}$ and $S_i^t$ is the state of spin $i$ at step $t$.

Synchronous dynamics is a standard deterministic dynamics for neural and
automata networks.
Unlike asynchronous dynamics, it does not in general guarantee a monotonic
decrease of the energy and may exhibit non-equilibrium behavior such as cycles
or periodic orbits~\cite{goles2013neural,coolen2001statistical}.
Therefore, the overlap obtained from this dynamics should not be interpreted as
an equilibrium order parameter or as a result of equilibrium Gibbs sampling.

Nevertheless, based on empirical observations in Hopfield-type associative
memory models, synchronous and asynchronous dynamics are often expected to give
closely related qualitative retrieval properties at the macroscopic
level~\cite{nishimori2001statistical}.
For this reason, we use the synchronous dynamics as a probe of the dynamical
stability and accessibility of the retrieval state, rather than as an
equilibrium sampling procedure.
This distinction between equilibrium analysis and retrieval dynamics is well
known in associative memory models
~\cite{amari1988statistical,okada1995hierarchy,gardner1987zero,eissfeller1992new}.

In the numerical simulations, we used system sizes
$N=1000,2000,3000,$ and $5000$.
For each parameter set, we generated $P=\alpha N$ random patterns
$\{\bm{\xi}^\mu\}_{\mu=1}^P$, with the retrieved pattern fixed as
$\bm{\xi}^1=(1,\ldots,1)$.
The Hebbian coupling matrix was constructed by Eq.~(\ref{eq:Hebb-rule}).
We then constructed the effective coupling matrix by Eq.~(\ref{eq:interaction-Jprime}).
The initial condition was chosen as
\begin{align}
\bm S^0=\bm \xi^1,
\end{align}
so that the initial overlap was $m_0=1$.
This initialization was chosen because the purpose of the simulations is not to
sample the equilibrium Gibbs measure, but to test whether the retrieved pattern
remains dynamically stable under the zero-temperature synchronous dynamics.
Starting from $\bm S^0=\bm\xi^1$ therefore probes the stability of the
retrieval state and the basin connected to the target pattern.
Consequently, the finite overlap obtained from this protocol should be
interpreted as a dynamical retrieval property, or a spinodal-like stability
measure, rather than as an equilibrium magnetization.

The maximum number of updates was $100$, and the iteration was stopped when
$\bm S^{t+1}=\bm S^t$ or when a period-two cycle was detected.
For a period-two cycle, the plotted overlap was averaged over the two
configurations in the cycle.
For runs that reached neither a fixed point nor a period-two cycle within
100 updates, the final overlap was estimated by averaging $m(t)$ over the
last 10 time steps.

\begin{figure}[t]
  \centering
  \includegraphics[width=0.8\linewidth]{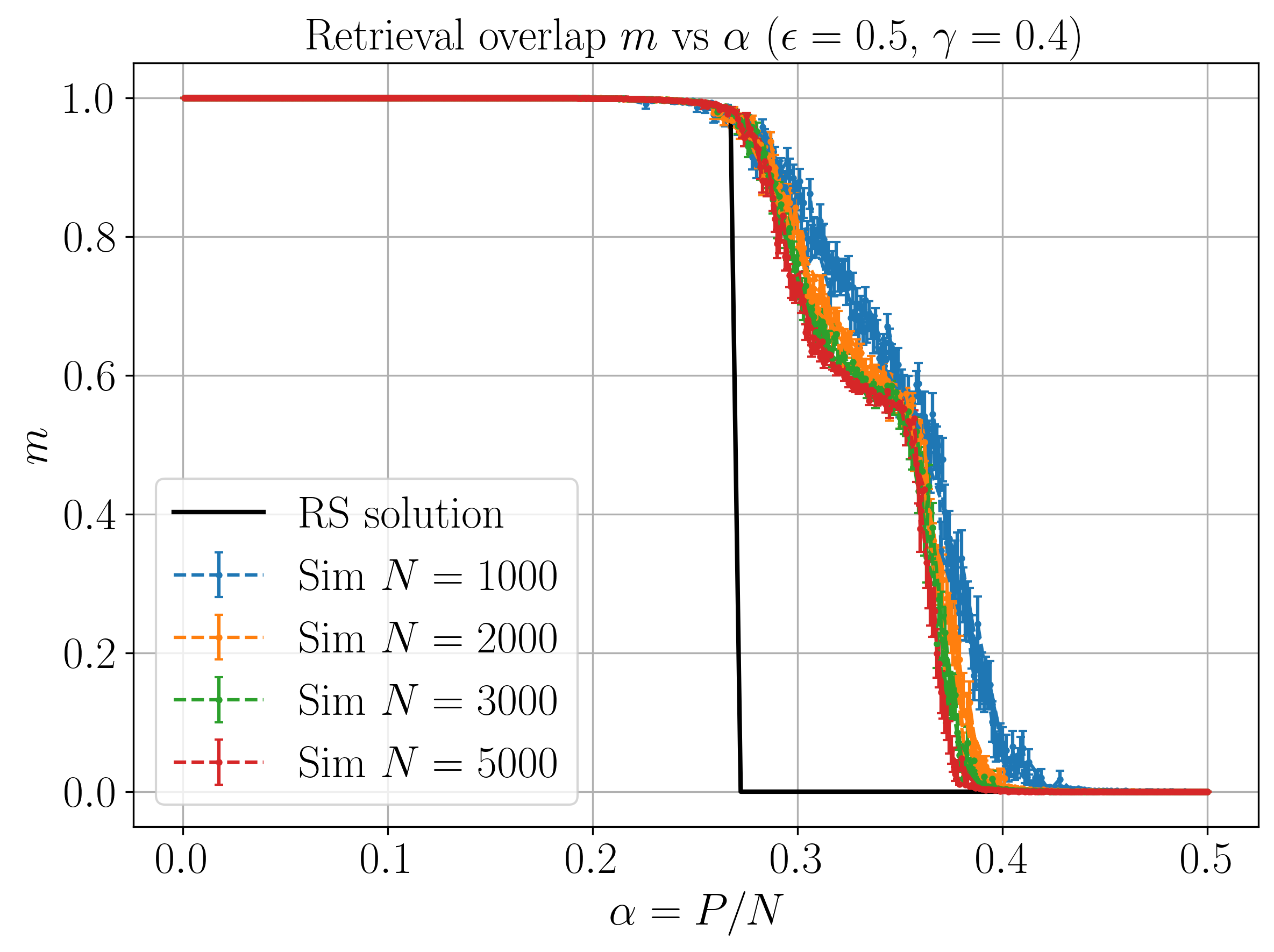}
    \caption{
    Overlap $m$ versus $\alpha$ at $\epsilon=0.5$ and $\gamma=0.4$.
    The pattern load was varied over $0.01\leq \alpha \leq 0.5$ with spacing $0.01$.
    The direct numerical simulations were performed for
    $N=1000,2000,3000,$ and $5000$, with 50 independent disorder realizations
    for each parameter value.
    The dynamics was initialized at the retrieved pattern, $\bm S^0=\bm\xi^1$,
    and the error bars represent the standard error over disorder realizations.
    Marker shapes summarize the outcomes among the 50 disorder realizations
    at each plotted point: circles indicate that at least $80\%$ of the realizations
    reached fixed points, triangles indicate that fewer than $80\%$ reached fixed
    points but at least $80\%$ reached either fixed points or period-two cycles,
    and crosses indicate that fewer than $80\%$ reached either fixed points or
    period-two cycles.
    The dotted line denotes the locally continued RS solution obtained from the
    saddle-point equations; its nonzero-$m$ part gives a spinodal-like estimate of
    the dynamical stability limit of the retrieved state.
    }
  \label{fig:m-vs-alpha-eps0.5-gamma0.4}
\end{figure}

\begin{figure}[t]
\centering
\begin{minipage}{0.48\linewidth}
\centering
\includegraphics[width=\linewidth]{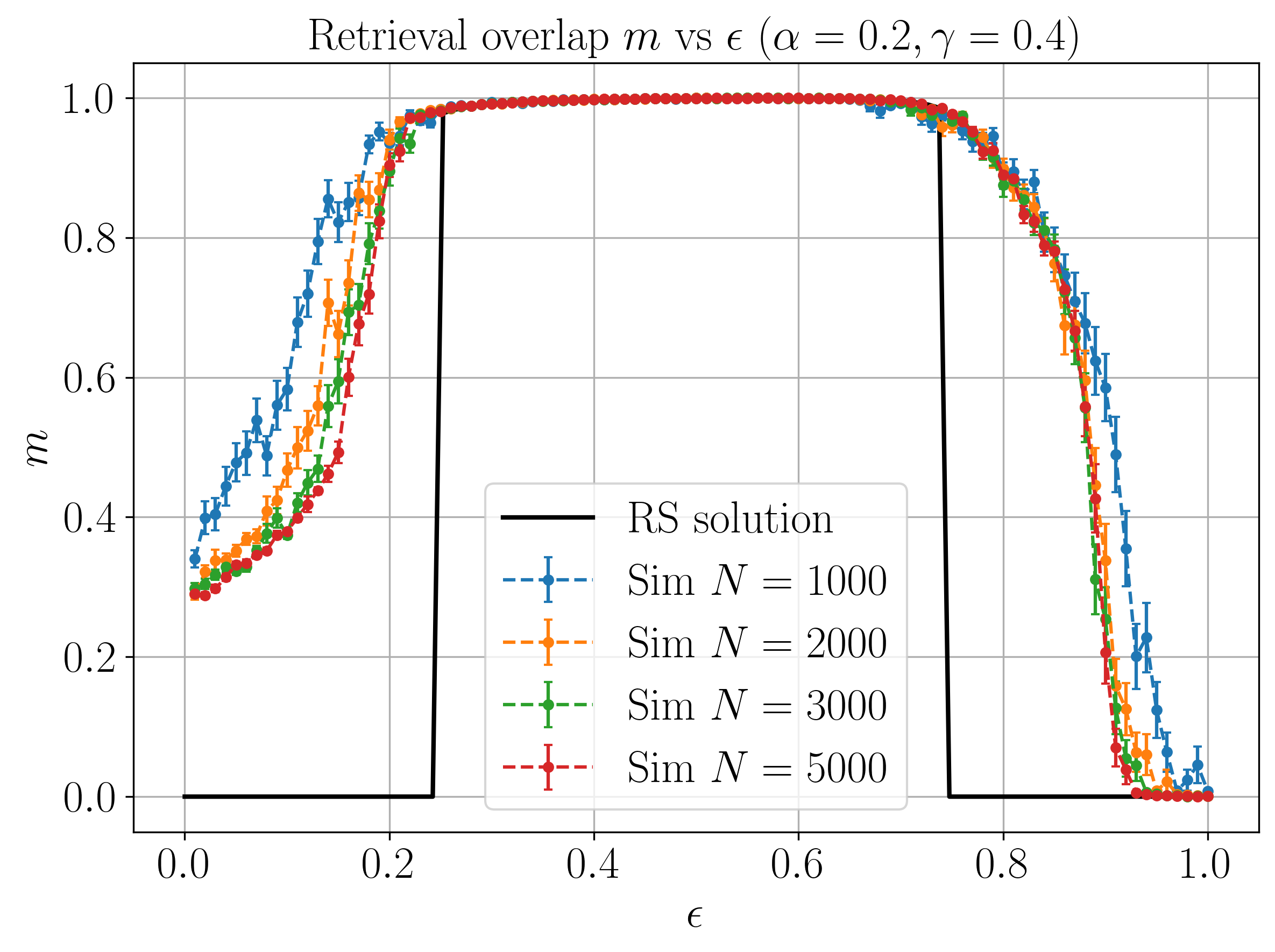}
\caption{
Overlap $m$ versus $\epsilon$ at $\alpha=0.2$ and $\gamma=0.4$.
The unlearning strength was varied over $0.01\leq \epsilon \leq 1.0$ with spacing $0.01$.
The numerical simulations were initialized at the retrieved pattern,
$\bm S^0=\bm\xi^1$.
The marker shapes follow the same convention as in
Fig.~\ref{fig:m-vs-alpha-eps0.5-gamma0.4}.
The dotted line has the same meaning as in
Fig.~\ref{fig:m-vs-alpha-eps0.5-gamma0.4}.
}
\label{fig:m-vs-eps-alpha0.2-gamma0.4}
\end{minipage}\hfill
\begin{minipage}{0.48\linewidth}
\centering
\includegraphics[width=\linewidth]{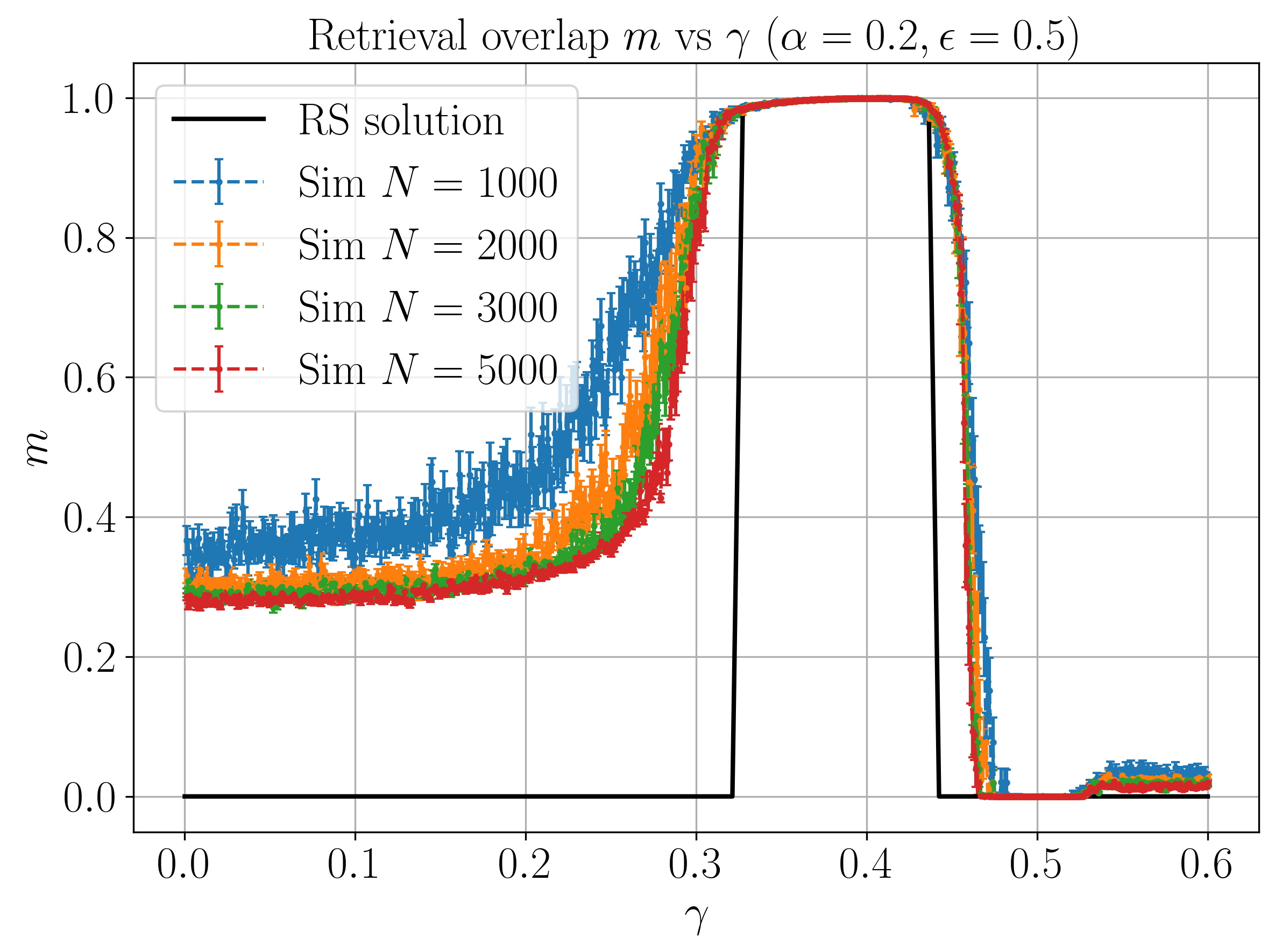}
\caption{
Overlap $m$ versus $\gamma$ at $\alpha=0.2$ and $\epsilon=0.5$.
The unlearning inverse temperature was varied over $0.01\leq \gamma \leq 0.6$ with spacing $0.01$.
The numerical simulations were initialized at the retrieved pattern,
$\bm S^0=\bm\xi^1$.
The marker shapes follow the same convention as in
Fig.~\ref{fig:m-vs-alpha-eps0.5-gamma0.4}.
The dotted line has the same meaning as in
Fig.~\ref{fig:m-vs-alpha-eps0.5-gamma0.4}.
}
\label{fig:m-vs-gamma-alpha0.2-eps0.5}
\end{minipage}
\end{figure}

Here,
\begin{align}
m(t)=\frac{1}{N}\sum_i \xi_i^1 S_i^t .
\end{align}
The plotted points and error bars represent the mean and standard error over
50 independent disorder realizations.
To compare the RS theory with this retrieval dynamics, we also followed the
locally continued RS solution of the saddle-point equations.
This branch was obtained by the ``warm start'' continuation: starting from a parameter point where
a solution with large $m$ exists, we used the solution at the previous
parameter value as the initial condition for solving the saddle-point equations
at the next parameter value.
In this procedure, the solution was followed locally, without selecting the
thermodynamically dominant solution by comparing free energies.
Thus, this branch should be distinguished from the thermodynamically stable RS
solution, which is selected by comparing the free energies of competing
solutions.
The endpoint of the nonzero-$m$ part of this continued solution is interpreted
as a spinodal point of the retrieval state.
The dotted lines in
Figs.~\ref{fig:m-vs-alpha-eps0.5-gamma0.4},
\ref{fig:m-vs-eps-alpha0.2-gamma0.4}, and
\ref{fig:m-vs-gamma-alpha0.2-eps0.5}
represent these locally continued RS solutions.

\FloatBarrier
Figure~\ref{fig:m-vs-alpha-eps0.5-gamma0.4} shows the dependence of $m$ on
$\alpha$ for $\epsilon=0.5$ and $\gamma=0.4$.
The thermodynamic RS analysis gives the estimate
$\alpha_{\mathrm{c}}(\epsilon=0.5,\gamma=0.4)\simeq 0.272$,
indicating that unlearning can improve retrieval performance within the RS
description.
In addition, the locally continued RS solution obtained from the saddle-point
equations is shown as a dotted line.
The numerical simulations initialized at $\bm S^0=\bm\xi^1$ are therefore more
naturally compared with this continued solution than with the thermodynamic
transition alone.

Above the thermodynamic transition point, the equilibrium RS solution is a
spin-glass state with vanishing retrieval overlap, $m=0$, while the spin-glass
order parameter remains finite, $q\neq0$.
However, the synchronous dynamics initialized at $\bm S^0=\bm\xi^1$ can still
show a finite overlap in this region.
This does not contradict the equilibrium RS prediction, because the simulation does not sample the equilibrium Gibbs measure.
Rather, the finite overlap reflects the persistence of a dynamically accessible
retrieval state, or a spinodal-like retrieval branch, under the chosen
initialization protocol, as commonly discussed in the context of retrieval
dynamics and metastable states in associative memory models
~\cite{amari1988statistical,okada1995hierarchy,gardner1987zero,eissfeller1992new}.
\FloatBarrier

Next, we fix $\alpha=0.2$ and vary the unlearning parameters $\epsilon$ and
$\gamma$.
Figures~\ref{fig:m-vs-eps-alpha0.2-gamma0.4} and
\ref{fig:m-vs-gamma-alpha0.2-eps0.5} show the dependence of the overlap $m$ on
$\epsilon$ and $\gamma$, respectively.
The dotted lines show the locally continued RS solutions obtained from the
saddle-point equations; their nonzero-$m$ parts may correspond to retrieval
branches associated with the initialization $\bm S^0=\bm\xi^1$.
In the high-overlap retrieval region, the numerical simulations initialized
from the retrieved pattern follow the RS solution reasonably well, supporting
the interpretation that the simulations probe the dynamical stability of the retrieval state rather than the thermodynamic equilibrium magnetization.
These results indicate that the retrieval performance is highly sensitive to
the choice of $\epsilon$ and $\gamma$.

The convergence behavior also depends on the overlap.
In regions where $m$ remains close to $1$, most runs reach fixed points or
period-two cycles within 100 updates, and the numerical results agree well with
the RS retrieval branch.
In regions where $m$ deviates substantially from $1$, the dynamics often does not reach either a fixed point or a period-two cycle within the same time window.
This does not affect the main interpretation, because the present simulations
are intended to test whether the retrieved pattern remains dynamically stable
from $\bm S^0=\bm\xi^1$, rather than to characterize the long-time dynamical state after retrieval has failed.

The disappearance of the continued retrieval branch indicates a spinodal-like
loss of the retrieval state, but it is not necessarily a
thermodynamic transition or a first-order transition.
A thermodynamic transition would be determined by comparing the free energies
of competing RS solutions.
In the present analysis, however, we focus only on the retrieval performance
measured by the overlap $m$, and do not further address the thermodynamic
properties of the competing branches.

We next discuss why the numerical overlap remains finite in
Figs.~\ref{fig:m-vs-eps-alpha0.2-gamma0.4} and
\ref{fig:m-vs-gamma-alpha0.2-eps0.5} even when $\epsilon\to0$ or $\gamma\to0$.
As discussed in Sec.~\ref{sec:model}, the effective interaction reduces to the
standard Hopfield coupling in the limits $\epsilon\to0$ or $\gamma\to0$,
up to the diagonal contribution that is removed in the present convention.
Therefore, for $\alpha=0.2$, the standard equilibrium theory of the Hopfield
model does not predict a thermodynamically stable retrieval state with finite
overlap.
However, the present simulations are initialized at the retrieved pattern,
$\bm S^0=\bm\xi^1$, and therefore probe the dynamical stability of the
retrieval state connected to this initial condition, rather than the equilibrium
Gibbs measure.
Moreover, as discussed above, synchronous dynamics can retain a finite overlap
associated with a dynamically accessible retrieval state or a locally continued
retrieval branch.
Thus, the finite overlap observed even when the unlearning parameters are
taken to $\epsilon\to0$ or $\gamma\to0$ should not be regarded as a
contradiction with the equilibrium Hopfield theory.
Rather, it reflects the stability of the retrieval branch reached from the
chosen initialization protocol.
This interpretation is consistent with the comparison to the continued
retrieval branch shown in
Fig.~\ref{fig:m-vs-gamma-alpha0.2-eps0.5}.

\subsection{Signal-to-Noise (SN) Ratio}

In Ref.~\cite{nokura1996unlearning}, Nokura analyzed the present model using a
qualitative argument based on a signal-to-noise (SN) analysis.
Here we briefly review this argument and then reformulate it within our
theoretical framework.

We first examine the stability of the state in which the spins are aligned exactly with
the first stored pattern, $\bm{S}=\bm{\xi}^1$.
In this situation, the local field acting on site $i$ is given by
\begin{align}
  h_i
  &= \sum_{j\neq i} J'_{ij} \xi_j^1 \notag \\
  \label{eq:local-field-Jprime}
  &= \sum_{j\neq i} J_{ij} \xi_j^1
  - \epsilon \sum_{j\neq i}
  \langle S_i S_j \rangle_\gamma \, \xi_j^1 .
\end{align}
From the update dynamics in Eq.~\eqref{eq:synchronous-update}, one expects that
the pattern $\bm{\xi}^1$ is stable if $\xi_i^1 h_i$ is positive for all, or at
least for almost all, sites.
To analyze this condition, the local field $h_i$ is decomposed into a signal
component, proportional to $\xi_i^1$, and a noise component that is uncorrelated
with $\xi_i^1$,
\begin{align}
  h_i = h_s \, \xi_i^1 + h_n .
\end{align}

Substituting the Hebbian coupling into
Eq.~\eqref{eq:local-field-Jprime} and extracting the terms proportional to
$\xi_i^1$ by performing a high-temperature expansion of the spin correlation
function, Nokura obtained the signal term
\begin{align}
  h_s = 1 - \frac{\epsilon\gamma}{1-\gamma}.
\end{align}
The magnitude of the noise term is estimated as
\begin{align}
  |h_n|
  \sim
  \sqrt{
    \left(
      1 - \frac{\epsilon\gamma}{(1-\gamma)^2}
    \right)^2 \alpha
    + \frac{\epsilon^2}{(1-\gamma)^2}
    \frac{A^2}{1-A}
  },
\end{align}
where $A=\alpha\gamma^2/(1-\gamma)^2$.
The ratio $(h_s/h_n)^2$ thus serves as an indicator of memory performance: when
it is large, the signal dominates over the noise and the pattern is stable,
whereas when it is small, retrieval fails.
This criterion provides a qualitative assessment of the effect of unlearning.

Following this approach, we examine the SN ratio within the present RS
framework based on the effective $J'$ model.
Since the effective interaction itself relies on the high-temperature
approximation for the spin correlation function, the comparison with
Ref.~\cite{nokura1996unlearning} should be understood as a qualitative
consistency check rather than as an exact equivalence.
The overlap $m$ is determined by
\begin{align}
  m
  = \int Dy \,
  \tanh \beta
  \left(
    \hat{m} - A\hat{u}
    + \frac{\delta}{\sqrt{2}}y
  \right),
\end{align}
where the argument of the hyperbolic tangent can be interpreted as an effective
field acting on a typical spin.
In direct analogy with Nokura's argument, we decompose this effective field into
a deterministic part and a fluctuating noise part.
The term $\delta y/\sqrt{2}$ represents the cross-talk noise generated by the
non-retrieved patterns, while the remaining contribution corresponds to the
systematic field associated with the retrieved pattern.

Using
\begin{align}
  \hat{u}
  = \gamma
  \frac{\sqrt{\epsilon} + \hat{\chi}_r}
  {1 + \alpha\gamma + 2\hat{\chi}_p - \gamma}\, m,
  \qquad
  \hat{m} = m,
\end{align}
the noise-free component of the effective field,
$\hat{m}-A\hat{u}$, plays the role of the signal $h_s$ in Nokura's terminology.
Similarly, the fluctuation amplitude $\delta/\sqrt{2}$ is identified with the
noise strength $h_n$.
We therefore define
\begin{align}
  h_s
  &= 1
  - A\gamma
  \frac{\sqrt{\epsilon} + \hat{\chi}_r}
  {1 + \alpha\gamma + 2\hat{\chi}_p - \gamma} \notag \\
  &= 1
  - \gamma
  \frac{(\hat{\chi}_r + \sqrt{\epsilon})^2}
  {(1 + \alpha\gamma + 2\hat{\chi}_p - \gamma)
   (1 + \alpha\gamma + 2\hat{\chi}_p)}, \\
  h_n
  &= \frac{\delta}{\sqrt{2}}
  = \sqrt{
  \frac{
    (2\hat{q} + \sqrt{\Delta_2} - A\hat{r})^2
    + (\hat{r} - 2A\hat{p} - A\sqrt{\Delta_2})^2
  }{
    2\hat{q} + 2\hat{p} + 2\sqrt{\Delta_2}
  }} .
\end{align}
This definition enables a direct and quantitative comparison with the SN-ratio
analysis of Nokura within a unified framework.

\begin{figure}[t]
  \centering
  \includegraphics[width=0.8\linewidth]{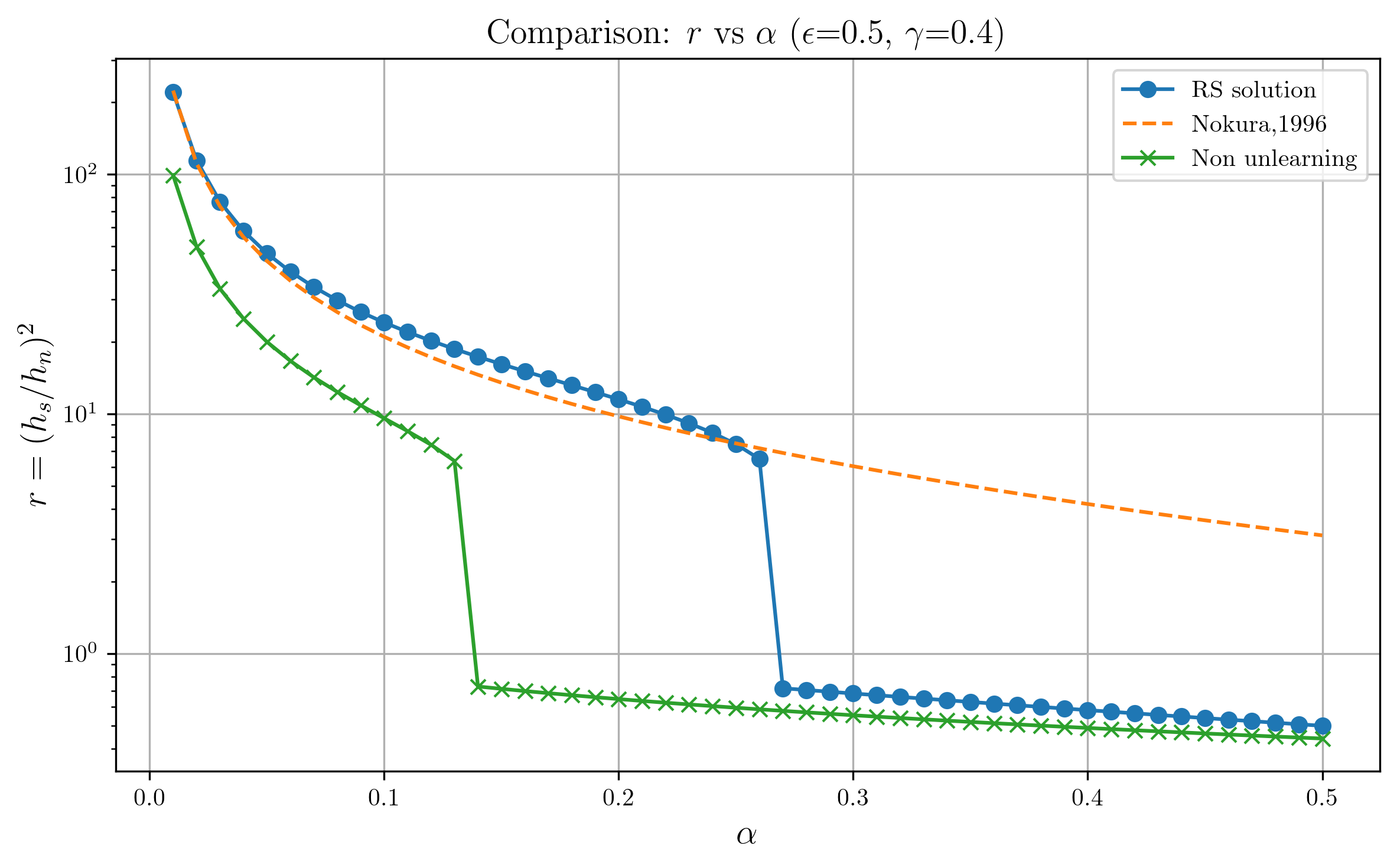}
  \caption{Signal-to-noise ratio at $\epsilon = 0.5$ and $\gamma = 0.4$.
  Blue: present theoretical result.
  Green: Hopfield model.
  Orange: SN-ratio analysis of Ref.~\cite{nokura1996unlearning}.}
  \label{fig:SN-ratio-eps0.5-gamma0.4}
\end{figure}

Figure~\ref{fig:SN-ratio-eps0.5-gamma0.4} shows that, within the present RS
description, the unlearning term increases the SN ratio compared with the
standard Hopfield model in the parameter region shown.
The result of the qualitative SN-ratio analysis of
Ref.~\cite{nokura1996unlearning} is also plotted for comparison.
The agreement in the retrieval region indicates that the present RS calculation
is qualitatively consistent with Nokura's argument under the high-temperature
approximation.
Thus, the SN-ratio comparison supports the interpretation that unlearning
suppresses noise components associated with non-retrieved patterns, although the
precise phase boundaries and capacities should still be interpreted as RS
estimates.

\subsection{Phase diagram and AT Line}
\label{subsec:phase-diagram-and-AT-line}
\begin{figure}[t]
  \centering
  \includegraphics[width=0.8\linewidth]{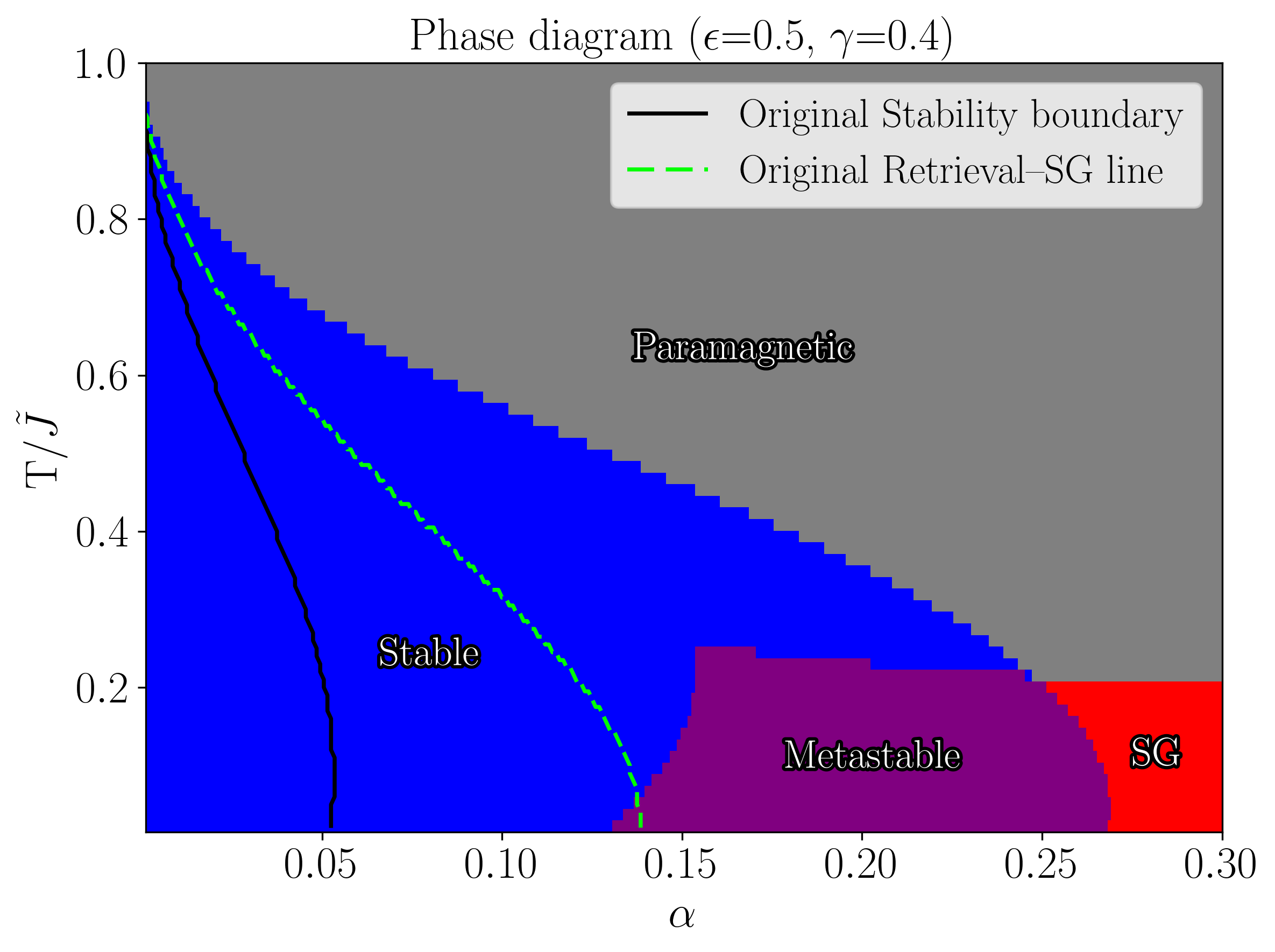}
  \caption{Phase diagram at $\epsilon = 0.5$ and $\gamma = 0.4$.
  The blue and purple regions correspond to the retrieval phase: blue indicates
  a stable retrieval state, while purple denotes a metastable one.
  The gray region represents the paramagnetic phase, and the red region the
  spin-glass (SG) phase.
  The constant $\tilde{J}$ is a scaling factor for the temperature.}
  \label{fig:phase-diagram-eps0.5-gamma0.4}
\end{figure}

Another advantage of the present approach over the previous 
SN
analysis is that it enables us to construct phase diagrams while explicitly
taking finite-temperature effects into account.
By numerically solving the saddle-point equations at finite temperature, we
obtain the finite-temperature phase diagram shown in
Fig.~\ref{fig:phase-diagram-eps0.5-gamma0.4}.

Unlearning reduces the overall magnitude of the effective interactions, which
results in a rescaling of the temperature axis in the phase diagram.
To account for this effect and to allow for a meaningful comparison on a common
vertical scale, we introduce a normalization constant $\tilde{J}$ and rescale
the temperature as $T/\tilde{J}$.
We define $\tilde{J}$ as
\begin{align}
\tilde{J} = \frac{1 - \epsilon\gamma - \gamma}{1 - \gamma}.
\end{align}
This expression is obtained by substituting the Hebbian interaction for a single
stored pattern into the definition of $J'$ in
Eq.~(\ref{eq:interaction-Jprime}), evaluating the resulting effective coupling,
and comparing it with the original Hopfield interaction $J$.

In Fig.~\ref{fig:phase-diagram-eps0.5-gamma0.4}, the black solid line and the green
dashed line indicate the reference phase boundaries of the Hopfield model
without unlearning.
The black line corresponds to the boundary between stable and metastable
retrieval states, while the green line denotes the boundary between the
retrieval and spin-glass phases.
Compared to these reference lines, both boundaries extend to larger values of
the pattern load $\alpha$ in the presence of unlearning.
Moreover, the spin-glass region is significantly reduced, indicating an
effective suppression of spurious memories.
On the other hand, as indicated by the purple region, there exists a finite
temperature above which the metastable part of the retrieval phase disappears.

This disappearance of the metastable retrieval phase at finite temperature does
not occur in the standard Hopfield model without unlearning.
At present, the detailed quantitative mechanism underlying this behavior remains unclear.
Nevertheless, a possible qualitative interpretation is related to the original
motivation of unlearning.
Unlearning was introduced to reduce the number or stability of unwanted
metastable states in the energy landscape.
In the present effective model, this mechanism may also weaken the metastable
retrieval basin at finite temperature, leading to the disappearance of the
locally continued retrieval branch.
A more detailed interpretation and a quantitative characterization of this
effect are left for future work.

To investigate the stability of the RS solution, we introduce a one-step replica symmetry breaking (1RSB) perturbation and evaluate the stability against small deviations. This procedure yields the de Almeida-Thouless (AT) condition~\cite{de1978stability}.
The details of the calculation are provided in
Appendix~\ref{sec:calculation-of-AT}; here we only summarize the result.
We define the following two matrices:
\begin{align}
U=
\begin{bmatrix}
\displaystyle \frac{\alpha}{2\Delta^2}(1+\beta\gamma\chi_p)^2 &
\displaystyle \frac{\alpha}{2\Delta^2}(-2\beta\gamma\chi_r)(1+\beta\gamma\chi_p) &
\displaystyle \frac{\alpha}{2\Delta^2}(\beta\gamma\chi_r)^2 \\[10pt]
\displaystyle \frac{\alpha\gamma}{\Delta^2}(-\beta\chi_r)(1+\beta\gamma\chi_p) &
\displaystyle \frac{\alpha\gamma}{\Delta^2}\!\left[-(1+\beta\gamma\chi_p)(1-\beta\chi_q)+(\beta\gamma\chi_r)^2\right] &
\displaystyle \frac{\alpha\gamma}{\Delta^2}(\beta\gamma\chi_r)(1-\beta\chi_q) \\[10pt]
\displaystyle \frac{\alpha\gamma^2}{2\Delta^2}(\beta\chi_r)^2 &
\displaystyle \frac{\alpha\gamma^2}{\Delta^2}(\beta\chi_r)(1-\beta\chi_q) &
\displaystyle \frac{\alpha\gamma^2}{2\Delta^2}(1-\beta\chi_q)^2
\end{bmatrix},
\end{align}
and
\begin{align}
V =
\begin{bmatrix}
2\beta^2\mathbb{E}_4 &
-2\beta^2\mathbb{E}_4\,A &
2\beta^2\mathbb{E}_4\,A^2 \\[10pt]
-2\beta^2\mathbb{E}_4\,A &
2\beta^2\mathbb{E}_4\,A^2 - \dfrac{\beta\chi_q}{1+\alpha\gamma+2\hat\chi_p} &
-2\beta^2\mathbb{E}_4\,A^3 + \dfrac{2\beta A\chi_q}{1+\alpha\gamma+2\hat\chi_p} \\[10pt]
2\beta^2\mathbb{E}_4\,A^2 &
-2\beta^2\mathbb{E}_4\,A^3 + \dfrac{2\beta A\chi_q}{1+\alpha\gamma+2\hat\chi_p} &
2\beta^2\mathbb{E}_4\,A^4 + \dfrac{2}{(1+\alpha\gamma+2\hat\chi_p)^2}
- \dfrac{4\beta A^{2}\chi_q}{1+\alpha\gamma+2\hat\chi_p}
\end{bmatrix}.
\end{align}
For compactness, we define
\begin{align}
\tilde H(y)
=
\hat m
-
A\hat u
+
\frac{\delta}{\sqrt{2}}y .
\end{align}
Then
\begin{align}
\mathbb{E}_4
\equiv
\int Dy\,
\left[
1-\tanh^2\!\left(
\beta \tilde H(y)
\right)
\right]^2 .
\end{align}
The RS solution is locally unstable if and only if the largest eigenvalue of
either $UV$ or $VU$ is greater than or equal to unity.
This criterion is the AT stability condition obtained from an infinitesimal
replica-symmetry-breaking perturbation around the RS saddle point
\cite{de1978stability}.

Strictly speaking, we do not perform a complete stability analysis against all
possible replica-symmetry-breaking fluctuations around the RS saddle point.
Such an analysis would require constructing the Hessian of the replicated free
energy in the full order-parameter space and examining its eigenmodes in all
replica directions, which is substantially more involved in the present model.
Instead, we consider the standard infinitesimal 1RSB-type perturbation around
the RS solution.
This perturbation can be regarded as probing the direction from the RS ansatz
toward a replica-symmetry-broken solution.
In the infinitesimal limit, this is reduced to the conventional construction that yields
the de Almeida--Thouless local stability condition for the RS saddle point
\cite{de1978stability}.
The resulting AT criterion is expressed above as the spectral-radius condition
for $UV$ or $VU$.

\begin{figure}[t]
  \centering
  \includegraphics[width=0.8\linewidth]{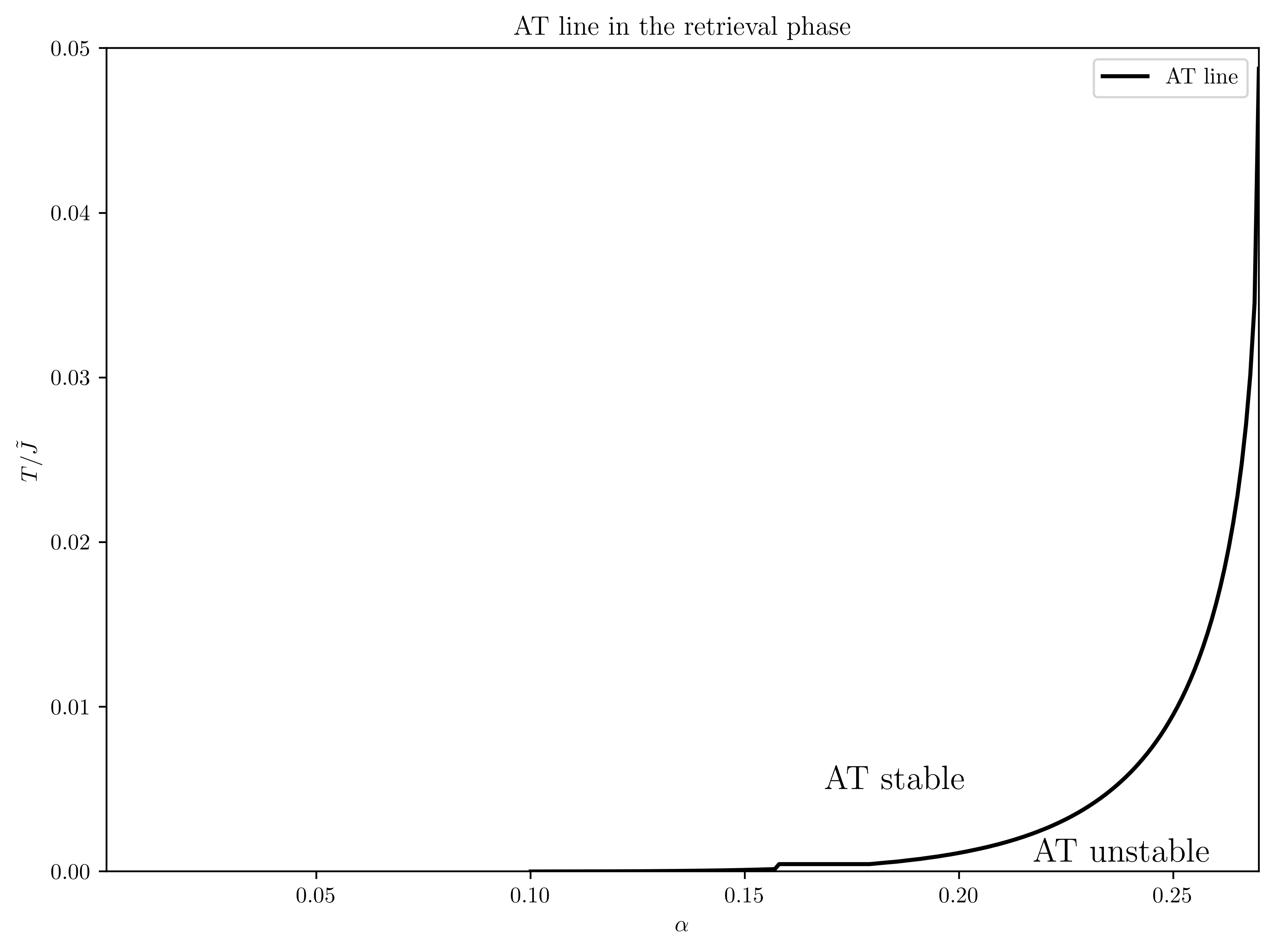}
    \caption{
    AT line at $\epsilon=0.5$ and $\gamma=0.4$, obtained from the largest
    eigenvalue of $UV$ or $VU$ evaluated at the RS saddle-point solution.
    The eigenvalue was evaluated in the temperature range $T\geq 10^{-4}$.
    }
  \label{fig:AT-line-eps0.5-gamma0.4}
\end{figure}

Figure~\ref{fig:AT-line-eps0.5-gamma0.4} shows the corresponding AT line.
For the parameter values shown here, the AT instability is mainly observed in
the very low-temperature region.
This behavior is reminiscent of the standard Hopfield model, where
replica-symmetry-breaking corrections are known to be important near saturation
and in the low-temperature regime
\cite{amit1985storing,crisanti1986saturation,amit1987statistical}.
Therefore, the zero-temperature capacity of the present $J'$ model should be
interpreted as an RS estimate, rather than as a fully RSB-stable capacity.
In particular, the stability near the reported RS capacity
$\alpha_{\mathrm{c}}\simeq0.272$ should be understood within this local AT
criterion, and possible RSB corrections may shift the precise capacity.
A complete classification of the low-temperature solution into a 1RSB or
full-RSB universality class is beyond the scope of the present work and remains
an important future problem.

\subsection{Heat Map of Memory Capacity}
\begin{figure}[t]
  \centering
  \includegraphics[width=0.8\linewidth]{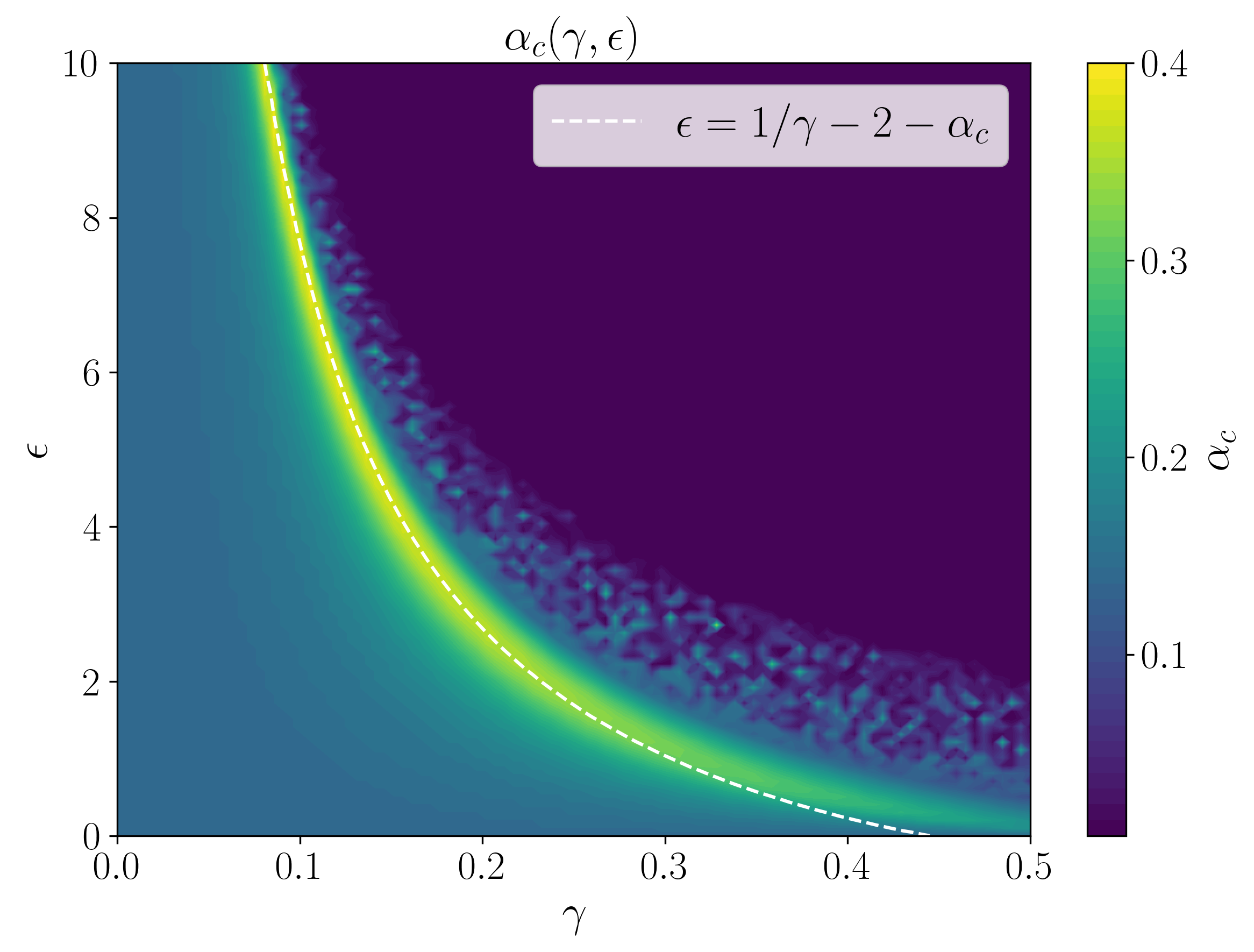}
    \caption{
    RS estimate of the memory capacity $\alpha_{\mathrm{c}}$ as a function of
    $\gamma$ and $\epsilon$.
    The white dashed curve represents
    $\epsilon=1/\gamma-2-0.4$, which is used as a reference curve motivated by
    the signal-to-noise optimization condition of
    Ref.~\cite{nokura1996unlearning}.
    }
  \label{fig:heatmap-of-alphac}
\end{figure}
In this subsection, we systematically investigate the RS estimate of the memory capacity $\alpha_{\mathrm{c}}$ as a function of the parameters $\epsilon$ and $\gamma$.
Figure~\ref{fig:heatmap-of-alphac} presents the results as a heatmap, where the horizontal and vertical axes represent $\gamma$ and $\epsilon$, respectively. We compute the RS estimate of the capacity $\alpha_{\mathrm{c}}$ at each point in this parameter space. As shown in the figure, there is a crescent-shaped region where the RS estimate of the capacity is
relatively large (indicated in yellow), suggesting that the performance of unlearning is highly sensitive to the choice of $\epsilon$ and $\gamma$. 

The white dashed curve is shown as a reference curve motivated by the
parameter relation obtained in the signal-to-noise analysis of
Ref.~\cite{nokura1996unlearning}.
In that analysis, the optimal parameters satisfy the approximate relation
\begin{align}
    \epsilon \simeq \frac{1}{\gamma}-2-\alpha .
\end{align}
In principle, when this relation is compared with the present capacity heat
map, $\alpha$ should be replaced by the RS estimate
$\alpha_{\mathrm{c}}(\gamma,\epsilon)$.
However, this leads to an implicit equation for $\epsilon$, since
$\alpha_{\mathrm{c}}$ itself depends on both $\gamma$ and $\epsilon$.
As a simple visual guide, we therefore plot
\begin{align}
    \epsilon = \frac{1}{\gamma}-2-0.4 ,
\end{align}
using the representative value $\alpha_{\mathrm{c}}\simeq0.4$, which is close
to the largest RS capacity values observed in the heat map.
The reference curve agrees reasonably well with the high-capacity region in the
small-$\gamma$ regime.
For larger $\gamma$, the curve deviates from the region of large
$\alpha_{\mathrm{c}}$.
This deviation is expected, because the relation
$\epsilon\simeq 1/\gamma-2-\alpha$ is derived as a small-$\gamma$
asymptotic condition, and because we have fixed $\alpha$ to the representative
value $0.4$ instead of solving the implicit equation with
$\alpha=\alpha_{\mathrm{c}}(\gamma,\epsilon)$.
Thus, in Fig.~\ref{fig:heatmap-of-alphac}, the white dashed curve should be
understood as a reference curve that captures the small-$\gamma$ trend of the
high-capacity region, rather than as an exact optimality condition valid over
the entire parameter space.

It is tempting to interpret this high-capacity region from a spectral viewpoint.
Indeed, at the level of the matrix expression in Eq.~(\ref{eq:interaction-Jprime}),
one may formally write a simple eigenvalue mapping.
If $Jv=\lambda v$, then
\begin{align}
    J'v
    =
    \left(
    \lambda-\frac{\epsilon}{1-\gamma\lambda}
    \right)v ,
\end{align}
and hence
$\lambda'=\lambda-\epsilon/(1-\gamma\lambda)$.
This expression suggests that the unlearning term deforms the effective
interaction in a mode-dependent manner.

However, deriving the optimal-parameter relation or the capacity phase diagram
directly from this spectral mapping is not straightforward.
In the retrieval problem, the target pattern must be treated separately as the
signal, while the remaining patterns contribute to cross-talk noise.
Moreover, the rank-one contribution of the target pattern also appears inside
the inverse term in $J'$.
Therefore, the simple mapping
$\lambda\mapsto\lambda-\epsilon/(1-\gamma\lambda)$
does not by itself determine the retrieval capacity.
For this reason, the high-capacity region and the reference curve in
Fig.~\ref{fig:heatmap-of-alphac} are interpreted here through the RS
saddle-point equations, rather than through the spectral mapping alone.

The same small-$\gamma$ condition can also be understood from the effective
interaction itself.
Using $\epsilon \simeq 1/\gamma-2-\alpha$ in the small-$\gamma$ expansion of
$J'$, one obtains
\begin{align}
    J' = \gamma \left( (2+\alpha)J-J^2 \right),
\end{align}
which reproduces the form discussed in Ref.~\cite{nokura1996unlearning}.
This expression suggests that the $J^2$ term plays a dominant role in the
unlearning process in this asymptotic regime.
The detailed replica calculation for this framework is provided in
Appendix~\ref{sec:replica-calculation-J2}.
\section{Conclusion}
\label{sec:conclusion}
In this paper, we presented a systematic statistical-mechanics analysis of a Hopfield
model incorporating unlearning based on spin correlations evaluated in the high-temperature
regime.
Focusing on the effective $J^\prime$ model proposed by Nokura and obtained
through the high-temperature approximation, we analyzed the model in the
thermodynamic limit using the replica method under the RS
ansatz and derived the corresponding free energy and self-consistent
saddle-point equations.

Our analysis provides a unified theoretical framework for evaluating the effects of unlearning
at both zero and finite temperatures.
Within this RS framework, we clarified how unlearning modifies the macroscopic
order parameters, enhances the signal-to-noise ratio within the RS description,
and can increase the RS estimate of the memory capacity by suppressing spurious
memories.
The resulting RS phase diagrams show that unlearning significantly reduces the
spin-glass region and expands the parameter range in which retrieval states are
stable.
These findings are consistent with, and extend beyond, the earlier qualitative
analysis of Nokura, providing a quantitative characterization of the RS estimate
of the memory capacity and its dependence on the unlearning parameters.

The theoretical predictions obtained from the replica analysis were shown to be
in reasonable agreement with numerical simulations in the retrieval phase.
This agreement should be interpreted as support for the dynamical stability of
the retrieval branch reached from the retrieved-pattern initialization, rather
than as a direct validation of equilibrium sampling.
In this sense, the numerical simulations probe retrieval dynamics and
spinodal-like behavior of the continued retrieval branch.
At finite temperatures, we further identified a metastable retrieval region
whose disappearance at higher temperatures is a distinctive feature of the
unlearning model and is absent in the standard Hopfield model.

Several open issues remain for future work.
A more precise analysis of the zero-temperature limit, including the effects of
replica-symmetry breaking, would be necessary to fully characterize the
spin-glass phase and the stability of retrieval states.
In particular, a 1RSB or full-RSB analysis would be required to determine
how RSB corrections modify the RS estimates of the phase boundaries and memory
capacity.
In addition, comparisons with true equilibrium sampling, for example by Monte
Carlo simulations, and with asynchronous dynamics would be important for
clarifying the relation between equilibrium properties and retrieval dynamics.
A dynamical analysis based on dynamical mean-field theory would also be useful
for understanding the behavior of synchronous update dynamics and its relation
to equilibrium predictions.
Finally, the validity and limitations of the high-temperature approximation used
to derive the effective $J^\prime$ model should be examined more systematically.
Extensions of the present framework to generalized associative memory models with higher-order
interactions, such as modern Hopfield-type models~\cite{krotov2016dense,ramsauer2021hopfield,demircigil2017model}, also constitute an interesting direction for further study.

\begin{acknowledgements}
This research was supported by Forefront Physics and Mathematics Program to Drive Transformation (FoPM), a World-leading Innovative Graduate Study (WINGS) Program, the University of Tokyo (ST), MEXT/JSPS KAKENHI Grant No. 22H05117 (YK), JSPS KAKENHI Grant Nos. 23K1690 (TT), 26K02981 (YK, TT),  and JST ACT-X Grant Number JPMJAX24CG (TT).

\end{acknowledgements}

\section*{Data availability statement}

The data and code that support the findings of this study are openly available
in Zenodo at \url{https://doi.org/10.5281/zenodo.20695795}.
The repository contains the processed numerical data, simulation and analysis
scripts, and plotting notebooks required to reproduce the figures.

\appendix

\section{Detailed Calculations of the Replica Method}
\label{detailed-calc-replica}

Before presenting the detailed calculation, we briefly summarize the structure of the derivation.
We first introduce auxiliary variables to rewrite the inverse interaction term and then replicate the partition function.
After separating the retrieved pattern from the remaining non-retrieved patterns, the latter are averaged by using Gaussian integral identities in the thermodynamic limit.
The resulting expression depends on the microscopic variables only through the macroscopic order parameters introduced in the main text.
Finally, we impose the RS parametrization and take the limit $n\to0$, which yields the free energy and the saddle-point equations.
This appendix provides the intermediate steps leading to the replicated free energy in Eq.~\eqref{eq:free-energy} and to the RS saddle-point equations given in Sec.~\ref{sec:analysis}.

To evaluate the last term in Eq.~\eqref{eq:average-of-partitionfunction}, we
introduce the variables
\begin{align}
  v^a = \frac{1}{\sqrt{N}}\sum_i \xi_i S_i^a,
  \qquad
  w^a = \frac{1}{\sqrt{N}}\sum_i \xi_i \phi_i^a .
\end{align}
Since $v^a$ and $w^a$ are sums of independent random variables, the central
limit theorem applies. Accordingly, $(v^a,w^a)$ are jointly Gaussian with
\begin{align}
  &\mathbb{E}\ab[v^a] = \mathbb{E}\ab[w^a] = 0, \\
  &\mathbb{E}\ab[v^a v^b] = \frac{1}{N}\sum_i S_i^a S_i^b \equiv q_{ab}, \\
  &\mathbb{E}\ab[v^a w^b] = \frac{1}{N}\sum_i S_i^a \phi_i^b \equiv r_{ab}, \\
  &\mathbb{E}\ab[w^a w^b] = \frac{1}{N}\sum_i \phi_i^a \phi_i^b \equiv p_{ab}.
\end{align}
These covariances can be summarized in the block form
\begin{align}
  \Sigma =
  \begin{pmatrix}
    Q & R \\
    R & P
  \end{pmatrix},
\end{align}
where $Q=(q_{ab})$, $R=(r_{ab})$, and $P=(p_{ab})$.
Replacing the average over the random patterns $\bm{\xi}$ by an average over
Gaussian variables $(\bm{v},\bm{w})\sim\mathcal{N}(0,\Sigma)$, we obtain
\begin{align}
  &\mathbb{E}_{\bm{\xi}}\ab[
    \exp\ab(
      \frac{\beta}{2N}\sum_{a}\ab(\sum_i \xi_i S_i^a)^2
      - \frac{\beta \gamma}{2N}\sum_{a}\ab(\sum_i \xi_i \phi_i^a)^2
    )
  ] \notag \\
  &\qquad=
  \mathbb{E}_{(\bm{v},\bm{w})\sim \mathcal{N}(0,\Sigma)}\ab[
    \exp\ab(
      \frac{\beta}{2}\sum_a (v^a)^2
      - \frac{\beta \gamma}{2}\sum_a (w^a)^2
    )
  ] .
\end{align}

We next introduce the macroscopic order parameters
\begin{align}
  m_a = \frac{1}{N}\sum_i S_i^a, \qquad
  u_a = \frac{1}{N}\sum_i \phi_i^a ,
\end{align}
and enforce their definitions, together with those of $q_{ab}$, $r_{ab}$, and
$p_{ab}$, by introducing conjugate variables via the Fourier representation of
the delta function (Eq.~\eqref{eq:fourier-representation-of-the-delta-function}).
This procedure leads to an integral representation of $\mathbb{E}[Z^n]$ of the
form
\begin{align}
  \mathbb{E}[Z^n]
  &= \int \exp\Bigg\{
  -N\Bigg(
    \beta\sum_a (\hat{m}_a m_a + \hat{u}_a u_a)
    + \beta \sum_{a,b}(\hat{q}_{ab}q_{ab} + \hat{r}_{ab}r_{ab} + \hat{p}_{ab}p_{ab})
    \notag \\
    &\hspace{7em}
    - \frac{\beta}{2}\sum_a m_a^2
    + \frac{\beta \gamma}{2}\sum_a u_a^2
  \Bigg)\Bigg\} \notag \\
  &\quad\times
  \ab(
    \mathbb{E}_{(\bm{v},\bm{w})\sim \mathcal{N}(0,\Sigma)}
    \ab[
      \exp\ab(
        \frac{\beta}{2}\sum_a (v^a)^2
        - \frac{\beta \gamma}{2}\sum_a (w^a)^2
      )
    ]
  )^{P-1} \notag \\
  &\quad\times
  \Tr \int_{-i\infty}^{i\infty}\prod_a \mathrm{d}\bm{\phi}^a\,
  \exp\Bigg(
    \beta \frac{1+\alpha\gamma}{2}\sum_{i,a}(\phi^a_i)^2
    + \beta \sqrt{\epsilon}\sum_{i,a}S_i^a \phi_i^a
    \notag \\
    &\hspace{7em}
    + \beta \sum_a \hat{m}_a \sum_i S_i^a
    + \beta \sum_a \hat{u}_a \sum_i \phi_i^a
    + \beta \sum_{a,b}\hat{q}_{ab}\sum_i S_i^a S_i^b
    \notag \\
    &\hspace{7em}
    + \beta \sum_{a,b}\hat{r}_{ab}\sum_i S_i^a \phi_i^b
    + \beta \sum_{a,b}\hat{p}_{ab}\sum_i \phi_i^a \phi_i^b
  \Bigg).
\end{align}
For brevity, we do not write the integrations over the order parameters and
their conjugates explicitly; in the thermodynamic limit they are evaluated by
the saddle-point method.

With these conjugate variables, the couplings between different sites are
removed, so that the trace factorizes over $i$.
The single-site contribution can be written as
\begin{align}
  L
  &= \frac{1 + \alpha \gamma}{2}\sum_a (\phi^a)^2
  + \sqrt{\epsilon}\sum_a S^a \phi^a
  + \sum_a \hat{m}_a S^a
  + \sum_a \hat{u}_a \phi^a
  + \sum_{a,b}\hat{q}_{ab} S^a S^b \notag \\
  &\quad
  + \sum_{a,b}\hat{r}_{ab} S^a \phi^b
  + \sum_{a,b}\hat{p}_{ab} \phi^a \phi^b ,
\end{align}
and hence the remaining contribution factorizes as
\begin{align}
  \left(
    \int_{-i\infty}^{i\infty} \prod_a \mathrm{d}\phi^a \,
    \Tr \, e^{\beta L}
  \right)^N .
\end{align}

We use the following Gaussian integral identity.
For an $n$-dimensional Gaussian vector $z$ with mean zero and covariance
matrix $\Sigma$, one has
\begin{align}
    \E_z\left[
    \exp\left(
    \frac{1}{2} z^{\mathsf T} B z
    \right)
    \right]
    =
    \det(I-\Sigma B)^{-1/2},
\end{align}

provided that the determinant is well defined and the integral is convergent.
Equivalently,
\begin{align}
\int \frac{\mathrm{d}^n x}{(2\pi)^{n/2}}
\exp\left(
-\frac{1}{2}x^{\mathsf T}Mx
\right)
=
\det M^{-1/2}
\end{align}
for a matrix $M$ for which the Gaussian integral is convergent.
In the following calculation, this identity is applied to the Gaussian variables
$(\bm v,\bm w)$ introduced above.
We now evaluate the Gaussian expectation over $(\bm{v},\bm{w})$ with covariance
matrix $\Sigma$.
Using the standard formula for Gaussian integrals, we obtain
\begin{align}
  &\mathbb{E}_{(\bm{v},\bm{w})\sim \mathcal{N}(0,\Sigma)}
  \ab[
    \exp\ab(
      \frac{\beta}{2}\sum_a (v^a)^2
      - \frac{\beta \gamma}{2}\sum_a (w^a)^2
    )
  ] \notag \\
  &= \ab(
    \det\ab[
      I - \Sigma
      \begin{pmatrix}
        \beta I & 0 \\
        0 & -\beta \gamma I
      \end{pmatrix}
    ]
  )^{-\frac{1}{2}},
\end{align}
where the determinant is understood by analytic continuation if necessary.
Rewriting the determinant yields
\begin{align}
  &=
  \ab(
    \det
    \begin{pmatrix}
      \frac{I}{\beta} - Q & -R \\
      -R & -\frac{I}{\beta \gamma} - P
    \end{pmatrix}
    \det
    \begin{pmatrix}
      \beta I & 0 \\
      0 & -\beta \gamma I
    \end{pmatrix}
  )^{-\frac{1}{2}} .
\end{align}
For the subsequent saddle-point analysis, it is convenient to apply an analytic
continuation and rewrite this as
\begin{align}
  &=
  \ab(
    \det
    \begin{pmatrix}
      \frac{iI}{\beta} - iQ & -iR \\
      -iR & -\frac{iI}{\beta \gamma} - iP
    \end{pmatrix}
    \det
    \begin{pmatrix}
      \beta I & 0 \\
      0 & \beta \gamma I
    \end{pmatrix}
  )^{-\frac{1}{2}} .
\end{align}
Taking the logarithm and dropping additive constants that do not affect the
saddle-point equations, we obtain
\begin{align}
  &\log
  \mathbb{E}_{(\bm{v},\bm{w})\sim \mathcal{N}(0,\Sigma)}
  \ab[
    \exp\ab(
      \frac{\beta}{2}\sum_a (v^a)^2
      - \frac{\beta \gamma}{2}\sum_a (w^a)^2
    )
  ] \notag \\
  &= -\frac{1}{2}
  \log
  \det
  \begin{pmatrix}
    \frac{iI}{\beta} - iQ & -iR \\
    -iR & -\frac{iI}{\beta \gamma} - iP
  \end{pmatrix}
  + \text{const}.
\end{align}
Using a Gaussian integral representation of the determinant, this can be
rewritten as
\begin{align}
  &= \log \int \mathrm{d}\bm{x}\,\mathrm{d}\bm{y}\,
  \exp\ab[
    -\frac{i\gamma}{2\beta}\sum_a (x^a)^2
    +\frac{i}{2\beta}\sum_a (y^a)^2
    +\frac{i\gamma}{2}\sum_{a,b}q_{ab}x^a x^b
    +\frac{i\gamma}{2}\sum_{a,b}p_{ab}y^a y^b
    + i\gamma\sum_{a,b}r_{ab}x^a y^b
  ] \notag \\
  &= \log \int \mathrm{d}\bm{x}\,\mathrm{d}\bm{y}\, e^{\beta K},
\end{align}
where $K$ denotes the quadratic form in the exponent; constant terms have been
absorbed into the normalization.

Collecting all contributions that scale extensively with $N$, we finally obtain
\begin{align}
  \mathbb{E}[Z^n]
  &= \int \exp\Bigg\{
  N\Bigg(
  -\beta\Bigl[
    \sum_a (\hat{m}_a m_a + \hat{u}_a u_a)
    + \sum_{a,b}(\hat{q}_{ab}q_{ab} + \hat{r}_{ab}r_{ab} + \hat{p}_{ab}p_{ab})
  \Bigr]
  + \frac{\beta}{2}\sum_a m_a^2
  - \frac{\beta \gamma}{2}\sum_a u_a^2 \notag \\
  &\hspace{8em}
  + \alpha \log \int \mathrm{d}\bm{x}\,\mathrm{d}\bm{y}\, e^{\beta K}
  + \log \int_{-i\infty}^{i\infty} \mathrm{d}\bm{\phi}\, \Tr e^{\beta L}
  \Bigg)
  \Bigg\}.
\end{align}
Here,
\begin{align}
  L
  &= \frac{1 + \alpha \gamma}{2}\sum_a (\phi^a)^2
  + \sqrt{\epsilon}\sum_a S^a \phi^a
  + \sum_a \hat{m}_a S^a
  + \sum_a \hat{u}_a \phi^a
  + \sum_{a,b}\hat{q}_{ab}S^a S^b \notag \\
  &\quad
  + \sum_{a,b}\hat{r}_{ab}S^a \phi^b
  + \sum_{a,b}\hat{p}_{ab}\phi^a \phi^b ,\\
  K
  &= -\frac{i\gamma}{2\beta}\sum_a (x^a)^2
  + \frac{i}{2\beta}\sum_a (y^a)^2
  + \frac{i\gamma}{2}\sum_{a,b}q_{ab}x^a x^b
  + \frac{i\gamma}{2}\sum_{a,b}p_{ab}y^a y^b
  + i\gamma\sum_{a,b}r_{ab}x^a y^b .
\end{align}

The saddle-point equations follow from extremizing the exponent with respect to
the order parameters and their conjugates, yielding
\begin{align}
  &\hat{m}_a = m_a, \qquad
  \hat{u}_a = -\gamma u_a, \\
  &m_a = \langle S^a\rangle_{L}, \qquad
  u_a = \langle \phi^a\rangle_{L}, \\
  &\hat{q}_{ab} = \frac{i\alpha \gamma}{2}\langle x^a x^b\rangle_{K}, \qquad
  \hat{r}_{ab} = i\alpha \gamma \langle x^a y^b\rangle_{K}, \qquad
  \hat{p}_{ab} = \frac{i\alpha \gamma}{2}\langle y^a y^b\rangle_{K}, \\
  &q_{ab} = \langle S^a S^b\rangle_{L}, \qquad
  r_{ab} = \langle S^a \phi^b\rangle_{L}, \qquad
  p_{ab} = \langle \phi^a \phi^b\rangle_{L},
\end{align}
where $\langle\cdot\rangle_{L}$ and $\langle\cdot\rangle_{K}$ denote averages
with weights $e^{\beta L}$ and $e^{\beta K}$, respectively.

At this stage, the saddle-point equations are still written for
replica-dependent order parameters.
Since the replicated action is invariant under permutations of the replica
indices, we next restrict the saddle point to the RS subspace introduced in
Sec.~\ref{sec:analysis}.
Substituting the RS parametrization into the replicated free energy and then
taking the limit $n\to0$, we obtain the RS free energy used in the main text.
The explicit RS saddle-point equations are obtained by differentiating this RS
free energy with respect to the order parameters and their conjugates.

For the RS reduction of the remaining Gaussian averages, it is convenient to
use the following standard property of Gaussian variables.
If $z_1$ and $z_2$ are independent standard normal variables, then
$a z_1 + b z_2$ is distributed as $\sqrt{a^2+b^2}\,z$, where $z$ is a standard
normal variable.
This relation allows us to reduce the relevant two-dimensional Gaussian average
to a one-dimensional one.

As a consistency check, the standard Hopfield equations are recovered when the
unlearning contribution is removed.
For $\epsilon\to0$, the coupling reduces to $J'\to J$, and the auxiliary
field associated with the inverse interaction term decouples from the spin
sector.
For $\gamma\to0$, one has
\begin{align}
(I-\gamma J)^{-1}\to I .
\end{align}
Since the diagonal part of $J'$ is ignored in the present formulation, the
off-diagonal interaction satisfies $J'_{ij}\to J_{ij}$ for $i\neq j$.
Thus, in both limits, the RS saddle-point equations reduce to those of the
standard Hopfield model.

Finally, in the zero-temperature limit $\beta\to\infty$, one has $q\to 1$ and
$\tanh(\beta \tilde{H})\to \mathrm{sgn}(\tilde{H})$, which leads to the
zero-temperature form of the saddle-point equations.

\section{Detailed calculation of AT line}
\label{sec:calculation-of-AT}
In this appendix, we derive the AT stability condition used in
Sec.~\ref{subsec:phase-diagram-and-AT-line}.
The calculation should be understood as a local stability analysis of the RS
saddle point, not as a full solution of the 1RSB or full-RSB variational
problem.
We introduce an infinitesimal 1RSB-type perturbation around the RS solution and
expand the saddle-point equations to first order in the deviations from the RS
values.
The resulting linearized equations determine whether the RS saddle point is
stable against replicon-like perturbations.

We consider a one-step replica-symmetry-breaking (1RSB) perturbation around the
RS saddle point. In the 1RSB ansatz, replica indices are grouped into blocks of
size $x\in[0,1]$ (the Parisi breaking parameter).
For each overlap, we introduce a subscript $1$ when two replicas belong to the
same block and a subscript $0$ when they belong to different blocks.
We then expand the saddle-point equations to first order in the deviations
between the 1RSB and RS solutions, and determine whether the RS solution is
stable against such perturbations. This procedure yields the
de Almeida--Thouless (AT) stability condition.

For a pair of replica indices $(a,b)$, there are three cases:
\begin{enumerate}
  \item $a\neq b$ and $a,b$ belong to the same block (e.g.\ $\beta \hat q_1$),
  \item $a\neq b$ and $a,b$ belong to different blocks (e.g.\ $\beta \hat q_0$),
  \item $a=b$ (e.g.\ $\beta \hat q_1 + \hat\chi_q$).
\end{enumerate}
As an illustration, we focus on the part associated with the term $K$.
After applying a Hubbard--Stratonovich transformation, we introduce auxiliary
Gaussian variables $\tilde{\bm z}$ and $\bm z$ and obtain an expression of the form
\begin{align}
  &\mathbb{E}_{\bm{z} \sim \mathcal{N}(0,\Omega_1)}
  \Bigg[
    \Bigg(
      \mathbb{E}_{\tilde{\bm z} \sim \mathcal{N}(0,\tilde{\Omega}_1)}
      \Bigg[
        \Bigg(
          \int \mathrm{d}y_1\,\mathrm{d}y_2\,
          \exp\Big(
            -\frac{i\gamma}{2}y_1^2 + \frac{i}{2}y_2^2
            + \frac{i\gamma \beta}{2}\chi_q y_1^2
            + \frac{i\gamma\beta}{2}\chi_p y_2^2  \notag \\
            &\hspace{12em}
            + i\gamma\beta \chi_r y_1 y_2
            + \sqrt{\beta\gamma}(z_1+\tilde z_1)y_1
            + \sqrt{\beta\gamma}(z_2+\tilde z_2)y_2
          \Big)
        \Bigg)^x
      \Bigg]
    \Bigg)^{n/x}
  \Bigg] \notag \\
  &\equiv
  \mathbb{E}_{\bm z}\!\left[
    \left(
      \mathbb{E}_{\tilde{\bm z}}\!\left[
        \left(
          \int \mathrm{d}\bm y\,
          \exp\Big(
            -\frac{1}{2}\bm y^{\mathsf T}G\bm y
            + \bm\Gamma^{\mathsf T}\bm y
          \Big)
        \right)^x
      \right]
    \right)^{n/x}
  \right],
\end{align}
where $G$ and $\bm\Gamma$ summarize the quadratic and linear terms in $\bm y$.

To extract the linear response with respect to the 1RSB perturbation, we need to
distinguish whether a pair of replicas $(a,b)$ belongs to the same block or to
different blocks. The relevant contributions can be written schematically as
follows. For case~(1) (same block, $a\neq b$),
\begin{align}
  \mathbb{E}_{\bm z}\!\Bigg[
  \Big(
    \mathbb{E}_{\tilde{\bm z}}\![F(\tilde{\bm z})^x]
  \Big)^{\frac{n}{x}-1}
  \Big(
    \mathbb{E}_{\tilde{\bm z}}\![F(\tilde{\bm z})^{x-2} (F_1(\tilde{\bm z}))^2]
  \Big)
  \Bigg],
\end{align}
whereas for case~(2) (different blocks, $a\neq b$),
\begin{align}
  \mathbb{E}_{\bm z}\!\Bigg[
  \Big(
    \mathbb{E}_{\tilde{\bm z}}\![F(\tilde{\bm z})^x]
  \Big)^{\frac{n}{x}-2}
  \Big(
    \mathbb{E}_{\tilde{\bm z}}\![F(\tilde{\bm z})^{x-1} F_1(\tilde{\bm z})]
  \Big)^2
  \Bigg],
\end{align}
with
\begin{align}
  F(\tilde{\bm z}) \equiv \int \mathrm{d}\bm y\, e^{(\sim)},
  \qquad
  F_1(\tilde{\bm z}) \equiv \int \mathrm{d}\bm y\, y_1\, e^{(\sim)} .
\end{align}

For fixed $\bm z$, it is convenient to regard the tilted measure
\begin{align}
  P(\tilde{\bm z})
  \equiv
  \frac{1}{\mathbb{E}_{\tilde{\bm z}}[F(\tilde{\bm z})^x]}\,
  \frac{F(\tilde{\bm z})^x}{\sqrt{(2\pi)^2\det\tilde\Omega_1}}\,
  \exp\!\left(
    -\frac{1}{2}\tilde{\bm z}^{\mathsf T}\tilde\Omega_1^{-1}\tilde{\bm z}
  \right)
\end{align}
as the effective distribution of $\tilde{\bm z}$.
Then the difference between case~(1) and case~(2) takes the form of a variance
under $P(\tilde{\bm z})$, which implies
\begin{align}
  \beta(\hat q_1-\hat q_0)
  = \frac{i\alpha\gamma}{2}\,
  \mathbb{E}_{\bm z}\!\left[
    \mathrm{Var}_{\tilde{\bm z}}
    \left[
      \frac{1}{\sqrt{\beta\gamma}}\,
      \pdv{}{\tilde z_1}\log F(\tilde{\bm z})
    \right]
  \right].
\end{align}
Similarly,
\begin{align}
  \beta(\hat p_1-\hat p_0)
  &= \frac{i\alpha\gamma}{2}\,
  \mathbb{E}_{\bm z}\!\left[
    \mathrm{Var}_{\tilde{\bm z}}
    \left[
      \frac{1}{\sqrt{\beta\gamma}}\,
      \pdv{}{\tilde z_2}\log F(\tilde{\bm z})
    \right]
  \right], \\
  \beta(\hat r_1-\hat r_0)
  &= i\alpha\gamma\,
  \mathbb{E}_{\bm z}\!\left[
    \mathrm{Cov}_{\tilde{\bm z}}
    \left[
      \frac{1}{\sqrt{\beta\gamma}}\,
      \pdv{}{\tilde z_1}\log F(\tilde{\bm z}),
      \frac{1}{\sqrt{\beta\gamma}}\,
      \pdv{}{\tilde z_2}\log F(\tilde{\bm z})
    \right]
  \right].
\end{align}

The Gaussian integral over $\bm y=(y_1,y_2)$ can be evaluated explicitly:
\begin{align}
  F(\tilde{\bm z})
  &= \int \mathrm{d}y_1\,\mathrm{d}y_2\,
  \exp\Big(
    -\frac{1}{2}\bm y^{\mathsf T}M\bm y
    + \sqrt{\beta\gamma}\,(z_1+\tilde z_1)y_1
    + \sqrt{\beta\gamma}\,(z_2+\tilde z_2)y_2
  \Big) \notag \\
  &= \Delta^{-1/2}
  \exp\!\left(
    \frac{iA}{2}(z_1+\tilde z_1)^2
    + \frac{iB}{2}(z_2+\tilde z_2)^2
    + iC(z_1+\tilde z_1)(z_2+\tilde z_2)
  \right),
\end{align}
where $\Delta$ is the determinant factor and the coefficients are
\begin{align}
  A &= -\frac{\beta}{\Delta}\,(1+\gamma\beta\chi_p), \\
  B &= \frac{\gamma\beta}{\Delta}\,(1-\beta\chi_q), \\
  C &= \frac{\gamma\beta^2}{\Delta}\,\chi_r .
\end{align}
Differentiating $\log F$ with respect to $\tilde z_1$ and $\tilde z_2$ gives
\begin{align}
  \pdv{}{\tilde z_1}\log F
  &= iA\tilde z_1 + iC\tilde z_2 + iAz_1 + iCz_2, \\
  \pdv{}{\tilde z_2}\log F
  &= iB\tilde z_2 + iC\tilde z_1 + iBz_2 + iCz_1 .
\end{align}
Since $z_1$ and $z_2$ drop out when taking $\mathrm{Var}_{\tilde{\bm z}}$,
we obtain a closed linear relation between the deviations,
\begin{align}
  \hat q_1-\hat q_0
  &=\frac{\alpha}{2\Delta^{2}}
  \Bigl[
    (1+\gamma\beta\chi_p)^{2}(q_1-q_0)
    +\gamma^{2}\beta^{2}\chi_r^{2}(p_1-p_0)
    -2\gamma\beta(1+\gamma\beta\chi_p)\chi_r(r_1-r_0)
  \Bigr], \\
  \hat p_1-\hat p_0
  &=\frac{\alpha\gamma^2}{2\Delta^{2}}
  \Bigl[
    \beta^{2}\chi_r^{2}(q_1-q_0)
    +(1-\beta\chi_q)^{2}(p_1-p_0)
    +2\beta(1-\beta\chi_q)\chi_r(r_1-r_0)
  \Bigr], \\
  \hat r_1-\hat r_0
  &=\frac{\alpha\gamma}{\Delta^{2}}
  \Bigl[
    -\beta(1+\gamma\beta\chi_p)\chi_r(q_1-q_0)
    +\gamma\beta(1-\beta\chi_q)\chi_r(p_1-p_0) \notag \\
    &\hspace{6em}
    +\bigl(-(1+\gamma\beta\chi_p)(1-\beta\chi_q)+\gamma\beta^{2}\chi_r^{2}\bigr)(r_1-r_0)
  \Bigr].
\end{align}

We now collect the deviations between the 1RSB and RS solutions into vectors
\begin{align*}
  \bm{v} \equiv
  \begin{bmatrix}
    q_1-q_0\\
    r_1-r_0\\
    p_1-p_0
  \end{bmatrix},
  \qquad
  \hat{\bm{v}} \equiv
  \begin{bmatrix}
    \hat q_1-\hat q_0\\
    \hat r_1-\hat r_0\\
    \hat p_1-\hat p_0
  \end{bmatrix}.
\end{align*}
The above relations can then be written compactly as
\begin{align}
  \hat{\bm v} = U\,\bm v .
\end{align}
Performing the same linearization for the single-site term $L$, we similarly
obtain
\begin{align}
  \bm v = V\,\hat{\bm v}.
\end{align}
Composing the two relations yields closed equations for $\bm v$ or
$\hat{\bm v}$.
Thus, the AT condition obtained here is a local criterion for the stability of
the RS solution.
If the largest eigenvalue of $UV$, equivalently of $VU$, is smaller than
unity, the RS solution is locally stable against the infinitesimal 1RSB
perturbation considered here.
If it is greater than or equal to unity, the RS solution is locally unstable.
This criterion is the one used in Sec.~\ref{subsec:phase-diagram-and-AT-line}
to assess the stability near the RS capacity
$\alpha_{\mathrm{c}}\simeq0.272$.

We emphasize that this analysis does not determine the stable RSB solution
beyond the instability line.
In particular, it does not determine the breaking point or the full
order-parameter function.
It only identifies the local instability of the RS saddle point.

\section{Replica Analysis at Optimal Parameters}
\label{sec:replica-calculation-J2}
\begin{figure}[t]
  \centering
  \includegraphics[width=0.8\linewidth]{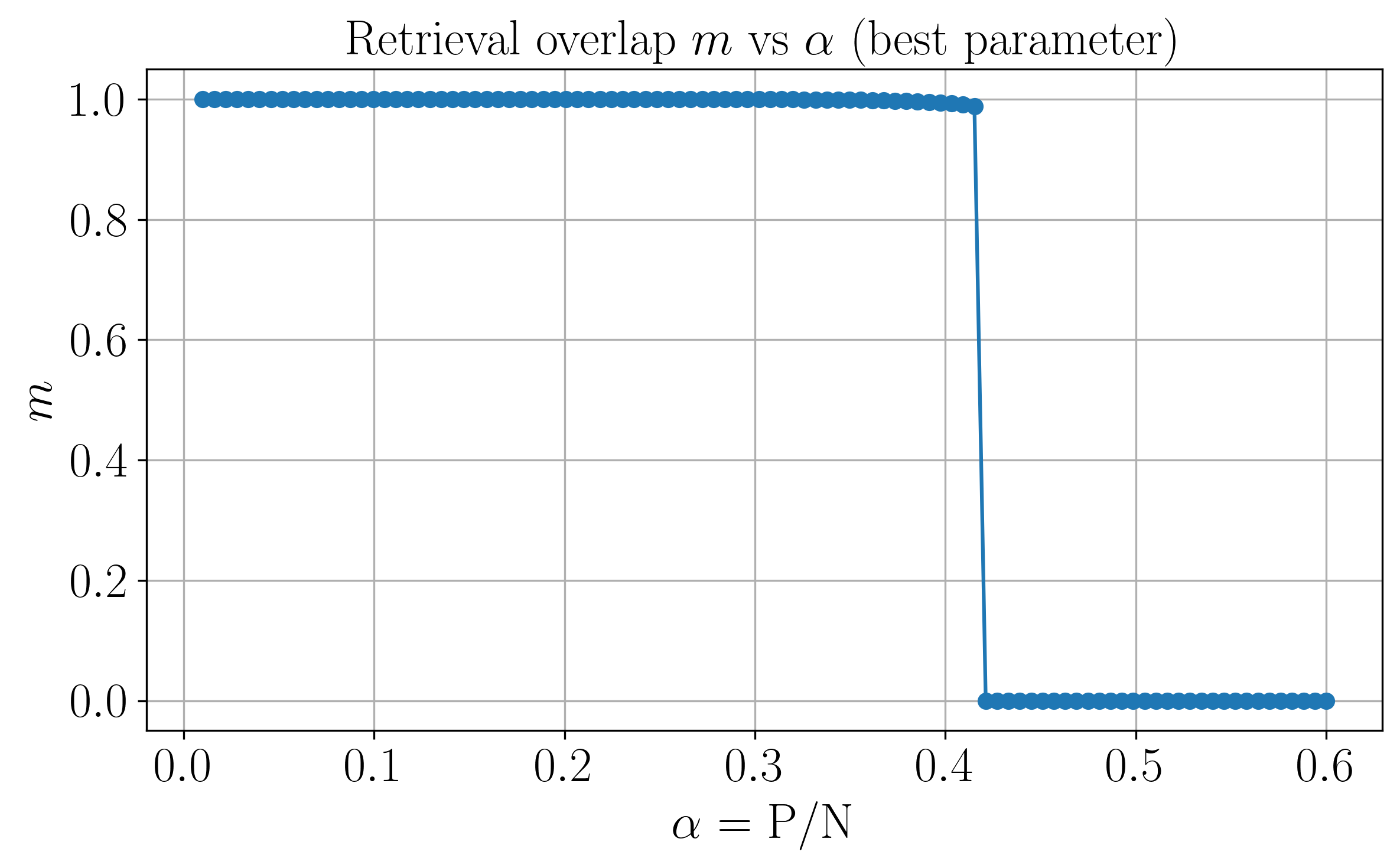}
  \caption{Overlap $m$ versus $\alpha$ in the small-$\gamma$ effective $J^2$ limit.}
  \label{fig:overlap-bestpara}
\end{figure}

We consider
the case of 
\begin{align}
  J' = \gamma \ab((2 + \alpha)J - J^2).
\end{align}
This yields
\begin{align}
  Z &= \Tr \exp\ab(\frac{\beta}{2}\bm{S}^{\mathsf{T}}J' \bm{S}) \notag \\
  &= \Tr \exp\ab(\frac{\beta \gamma (2 + \alpha)}{2}\bm{S}^{\mathsf{T}}J\bm{S} - \frac{\beta \gamma}{2} \bm{S}^{\mathsf{T}}J^2 \bm{S}).
\end{align}
Here,
\begin{align}
  \Tr \exp\ab(-\frac{\beta\gamma}{2}\bm{S}^{\mathsf{T}}J^2 \bm{S}) &= \Tr \exp\ab(-\frac{\beta\gamma}{2}\ab(J \bm{S})^2) \notag \\
  &= \int \prod_i \d \phi_i \Tr \exp\ab(-\frac{\beta\gamma}{2}\sum_i \phi_i^2 + i\beta\gamma\bm{\phi}^{\mathsf{T}}J\bm{S}) \notag \\
  &= \int_{-i\infty}^{i\infty} \prod_i \d \phi_i \Tr \exp\ab(\frac{\beta \gamma}{2}\sum_i \phi_i^2 + \beta\gamma\bm{\phi}^{\mathsf{T}}J\bm{S}).
\end{align}
thus introducing auxiliary variables. Proceeding as in Appendix~\ref{detailed-calc-replica}, we obtain
\begin{align}
  &m = \mathbb{E}_{\bm{z}\sim \mathcal{N}(0,\Omega_2)} \ab[ \tanh \beta \gamma \ab(\hat{m} + z_1 -{\frac{\left(\hat{r}_1 -  \alpha  \right) \left(\hat{u} + z_{2}\right)}{2 \hat{p}_1 + 1}})] \\
  &q_0 = \mathbb{E}_{\bm{z}\sim \mathcal{N}(0,\Omega_2)} \ab[ \tanh^2 \beta \gamma \ab(\hat{m} + z_1 -{\frac{\left(\hat{r}_1 -  \alpha \right) \left(\hat{u} + z_{2}\right)}{2 \hat{p}_1 + 1}})] \\
  &q_1 = 1 - q_0 \\
  &u = -\frac{1}{2\hat{p}_1 + 1} \ab((- \alpha + \hat{r}_1)\mathbb{E}\ab[\tanh] + \hat{u}) \\
  &p_1 = - \frac{1}{\beta \gamma (2\hat{p}_1 + 1)} + \frac{\left(\hat{r}_1 -  \alpha \right)^2\ab(1 - \mathbb{E}\ab[\tanh^2])}{ (2\hat{p}_1 + 1)^2}  \\
  &p_0 = \frac{1}{(2\hat{p}_1 + 1)^2}\ab((- \alpha + \hat{r}_1)^2\mathbb{E}\ab[\tanh^2] + 2(- \alpha + \hat{r}_1)(\hat{u}\mathbb{E}\ab[\tanh] + \mathbb{E}\ab[z_{2}\tanh]) + \hat{u}^2 + 2\hat{p}_0) \\
  &r_1 = -\frac{1}{2\hat{p}_1 + 1}\ab(-\alpha + \hat{r}_1)\ab(1 - \mathbb{E}\ab[\tanh^2]) \\
  &r_0 = -\frac{1}{2\hat{p}_1 + 1}\ab((- \alpha + \hat{r}_1)\mathbb{E}\ab[\tanh^2] + \hat{u}\mathbb{E}\ab[\tanh] + \mathbb{E}\ab[z_2 \tanh]).
\end{align}
\begin{align}
  \hat{m} &= (2 + \alpha)m + u \\
  \hat{u} &= m \\
  \hat{q}_1 &= \frac{\alpha}{2\Delta} \ab(\beta \gamma p_1 + 2 + \alpha)  \\
  \hat{q}_0 &= \frac{\alpha}{2\Delta^2}\ab(\ab(\beta \gamma p_1 + 2 + \alpha)^2 q_0 + 2 \ab(\beta \gamma p_1 + 2 + \alpha)\ab(1 - \beta \gamma r_1) r_0 + \ab(1 - \beta \gamma r_1)^2 p_0) \\
  \hat{p}_1 &= \frac{\alpha \beta \gamma}{2\Delta}q_1  \notag \\
  \hat{p}_0 &= \frac{\alpha}{2\Delta^2}\ab(\ab(1 - \beta \gamma r_1)^2  q_0 + 2\beta \gamma q_1 \ab(1 - \beta \gamma r_1)r_0 + \gamma^2 \beta^2 q_1^2  p_0) \\
  \hat{r}_1 &= \frac{\alpha}{\Delta}\ab(1 - \beta \gamma r_1)  \\
  \hat{r}_0 &= \frac{\alpha}{\Delta^2}\Bigg(\ab(\beta \gamma p_1 + 2 + \alpha)\ab(1 - \gamma \beta r_1) q_0 + \beta \gamma q_1 \ab(\beta \gamma p_1 + 2 + \alpha)r_0 \notag \\
  & + \ab(1 - \gamma \beta r_1)^2 r_0 + \beta \gamma q_1 \ab(1 - \gamma \beta r_1)p_0 \Bigg).
\end{align}
are obtained.
By taking $\beta \gamma \rightarrow \infty$ so that $\beta\gamma q_1 = C, \beta\gamma p_1 = E, \beta\gamma r_1 = F$,
\begin{align}
  &m = \mathrm{erf}\ab(\frac{\hat{m} - A\hat{u}}{\delta}) \\
  &C = \frac{2}{\delta\sqrt{\pi}} \exp \ab(-\frac{(\hat{m} - A\hat{u})^2}{\delta^2}) \\
  &u = -\frac{1}{2\hat{p}_1 + 1} \ab((-\alpha + \hat{r}_1)m + \hat{u}) \\
  &E = - \frac{1}{2\hat{p}_1 + 1} + \frac{\left(\hat{r}_1 -  \alpha \right)^2C}{ (2\hat{p}_1 + 1)^2}  \\
  &p_0 = \frac{1}{(2\hat{p}_1 + 1)^2}\ab((- \alpha  + \hat{r}_1)^2 + 2(- \alpha  + \hat{r}_1)(\hat{u}m + C\ab(\hat{r}_0 - 2\hat{p}_0 A)) + \hat{u}^2 + 2\hat{p}_0) \\
  &F = -\frac{1}{2\hat{p}_1 + 1}\ab(-\alpha + \hat{r}_1)C \\
  &r_0 = -\frac{1}{2\hat{p}_1 + 1}\ab(-\alpha + \hat{r}_1 + \hat{u}m + C\ab(\hat{r}_0 - 2\hat{p}_0 A)).
\end{align}
\begin{align}
  \hat{m} &= (2 + \alpha)m + u \\
  \hat{u} &= m \\
  \hat{q}_1 &= \frac{\alpha}{2\Delta} \ab( E + 2 + \alpha) \\
  \hat{q}_0 &= \frac{\alpha}{2\Delta^2}\ab(\ab(E + 2 + \alpha)^2 + 2 \ab(E + 2 + \alpha)\ab(1 - F) r_0 + \ab(1 - F)^2 p_0) \\
  \hat{p}_1 &= \frac{\alpha}{2\Delta}C \\
  \hat{p}_0 &= \frac{\alpha}{2\Delta^2}\ab(\ab(1 - F)^2 + 2C \ab(1 - F) r_0 + C^2  p_0) \\
  \hat{r}_1 &= \frac{\alpha}{\Delta}\ab(1 - F)  \\
  \hat{r}_0 &= \frac{\alpha}{\Delta^2}\Bigg(\ab(E + 2 + \alpha)\ab(1 - F) + C \ab(E + 2 + \alpha) r_0 \notag \\
  & + \ab(1 - F)^2 r_0 + C \ab(1 - F) p_0 \Bigg).
\end{align}
Also, $A,\Delta,\delta,\Delta_2$ are
\begin{align}
  &A = \frac{\hat{r}_1 - \alpha}{2\hat{p}_1 + 1} \\
  &\Delta = -C \ab(E + 2 + \alpha) + \ab(1- F)^2 \\
  &\delta = \sqrt{\frac{(2\hat{q}_0 + \sqrt{\Delta_2} - A\hat{r}_0)^2 + (\hat{r}_0 - 2A\hat{p}_0 - A\sqrt{\Delta_2})^2}{\hat{q}_0 + \hat{p}_0 + \sqrt{\Delta_2}}} \\
  &\Delta_2 = 4\hat{q}_0 \hat{p}_0 - \hat{r}_0^2.
\end{align}
Using these saddle-point equations to plot the overlap $m$, the result is as shown in Fig.~\ref{fig:overlap-bestpara}.
The RS estimate of the memory capacity is approximately
$\alpha_{\mathrm{c}}\simeq0.42$, which is consistent with the value reported
from the signal-to-noise analysis of Ref.~\cite{nokura1996unlearning}.

\end{document}